\documentclass[aps,prd,longbibliography ,twocolumn,nofootinbib,10pt]{revtex4-2}
\date{\today}

\usepackage{amsmath}
\usepackage{amsfonts}
\usepackage{graphicx}
\usepackage{xcolor}
\usepackage{dsfont}
\usepackage[german, english]{babel}
\usepackage{amssymb}
\usepackage{ifthen}
\usepackage{soul}
\usepackage{color}
\usepackage{xcolor}

\usepackage{bm}

\newcommand{\appref}[1]{Appendix \ref{#1}}
\newcommand{\figref}[1]{Fig. \ref{#1}}
\renewcommand{\eqref}[1]{Eq. (\ref{#1})}
\newcommand{\ii}{\mathrm{i}}

\RequirePackage[colorlinks=true,allcolors=blue,
unicode=true,hypertexnames=false]{hyperref}
\hypersetup{pdfstartview={XYZ null null 1.25}}

\graphicspath{{images/}}

\begin{document}

\author{Albert Samoilenka}
\email{albsam@kth.se}
\author{Egor Babaev}
\affiliation{Department of Physics, KTH-Royal Institute of Technology, SE-10691, Stockholm, Sweden}

\title{Ground state fractal crystals}

\begin{abstract}
We propose a generalization of the crystalline order: the ground state fractal crystal.
We demonstrate that by deriving a simple continuous-space-discrete-field (CSDF) model whose ground state is a crystal where each unit cell is a fractal.
\end{abstract}

\maketitle

\section{Introduction}
Matter, described by quantum fields in a continuous space, can spontaneously break space translation symmetry by self-organizing into a periodic structure.
This phenomenon of crystallization is one of the cornerstone concepts in physics.
Crystals realize various states of condensed matter, such as metals, insulators, and superconductors.
In recent decades multiple discussions focused on the generalizations of the crystallization phenomenon.
One such concept is the moire crystals in twisted multi-layer materials \cite{cao2018unconventional}, leading to crystals with extremely large unit cells.
A separate class of order is quasicrystals \cite{shechtman1984metallic,levine1984quasicrystals}.
Another generalization that attracted interest for a long time is the class of order where a classical field demonstrates crystallization coexisting with the spontaneous breaking of additional symmetries.
The most known state of a crystal in a classical field is a vortex lattice in a superconductor.
This is also the case in a class of supersolids, i.e. the systems that break transition symmetry and have superfluid order (for a review see \cite{Svistunov2015}).
A class of systems, the so-called Fulde-Ferrell-Larkin-Ovchinnikov superconductors \cite{LarkinOvchinnikov1964} exhibit a crystallization in the form of superconducting Cooper-pair-density-wave.
Corresponding transition in the classical field theory is a particular case of the Lifshitz point.
It has been argued that dense quark matter in the cores of neutron stars is such a crystal \cite{alford2001crystalline}.
Cluster crystal is another different type of crystallization where a unit cell is a cluster of particles \cite{malescio2003stripe}.
This state is believed to form in outer regions of neutron stars, and has direct counterparts in soft matter \cite{caplan2017colloquium} and the quantum Hall effect \cite{fogler1996ground,shapere2012classical}.
Another topic of recent interest is time crystals: where the crystallization is in time or in time and space \cite{wilczek2012quantum,shapere2012classical}.

In this work, we propose a new generalization of the concept of a crystal: the ground state fractal crystal.

Fractals are ubiquitous in nature. 
In condensed matter, they usually appear as a result of a dynamic/kinetic process \cite{liu1986fractals,Nakayama2009}.
These random fractals were found, for example, in liquid crystal colloids by self-assembly \cite{solodkov2019self} and in polydisperse emulsions \cite{kwok2020apollonian}.
Moreover, deterministic fractals can appear as boundary phenomena arising due to competing effects in bulk and at interfaces.
An example of the latter is the Landau pattern in type-I superconductors \cite{landau1938intermediate} or states similar to Apollonian packing of circles in smectic A liquid crystals \cite{meyer2009focal}.
Fractal solitonic excitations were discussed in models such as the Davey-Stewartson model \cite{fractal_solitons}.

Below we investigate the possibility of a different state the ground state fractal crystal.
We define it as a state that satisfies the following conditions:

\begin{itemize}
\item The state should spontaneously break space translation symmetry down to a crystalline group
\item The unit cell of the resulting crystal should have an infinite number of elements, with each unit cell forming a fractal
\item The state should present an energy minimum of a Hamiltonian that respects translation invariance.
\end{itemize}

A particularly interesting question is whether there are classical field theories with such a ground state.
Classical field theories could be grouped into four categories by continuous/discrete space and field.
For example, the Ising model \cite{ising1925contribution} is a discrete space, discrete field model.
While the lattice XY model is a discrete space continuous field model.
An example of a continuous space continuous field model is Ginzburg-Landau theory \cite{Ginzburg1950}.
All these models have a uniform or crystal-like modulated ground states.
To realize a fractal crystal that does not suffer short-distance cutoff, only models with continuous space can be potential candidates.

\section{CSDF model derivation}
Here we formulate the simplest continuum-space-discrete-field (CSDF) model\footnote{This model is in some sense related to elasticity theory \cite{surface_elasticity_theory,interface_elasticity_theory,chhapadia2011curvature,javili2018aspects} which also studies zero-thickness limit of interfaces, which are obtained from homogenization of microscopic models. At the same time our model is rather different and is dependent only on curvature and its derivatives and is derived from different generic considerations.} that can have fractal crystal as a ground state.
To phenomenologically derive the CSDF model, an analogy with  
static Cahn-Hilliard \cite{CahnHilliard} model is useful.
So, firstly let us briefly recap the phenomenological derivation of the static Cahn-Hilliard model that describes structure formation in the  standard problem of phase separation \cite{CahnHilliard}:
\begin{equation}\label{CH}
\begin{gathered}
F_{CH}[c(\textbf{r})] = \int f\left( c(\textbf{r}), \nabla c(\textbf{r}), \nabla^2 c(\textbf{r}), .. \right) d\textbf{r} \\
f = V(c) + \gamma (\nabla c)^2 + ..
\end{gathered}
\end{equation}

where space is two dimensional $\textbf{r} = (x, y)$, the  usually  used form of the potential is $V(c) = - 2 c^2 + c^4$, and $c$ is order parameter of the model. The model describes a binary system.
We will refer to $c = 1$ as the first phase and $c = -1$ as the second phase (in general phases can have different values of $c$).
Note, that to justify expansion in orders of $c$ and derivatives in the Cahn-Hilliard model one assumes that $c$ is small and slowly changing in space.

\begin{figure}
\centering
\includegraphics[width=0.99\linewidth]{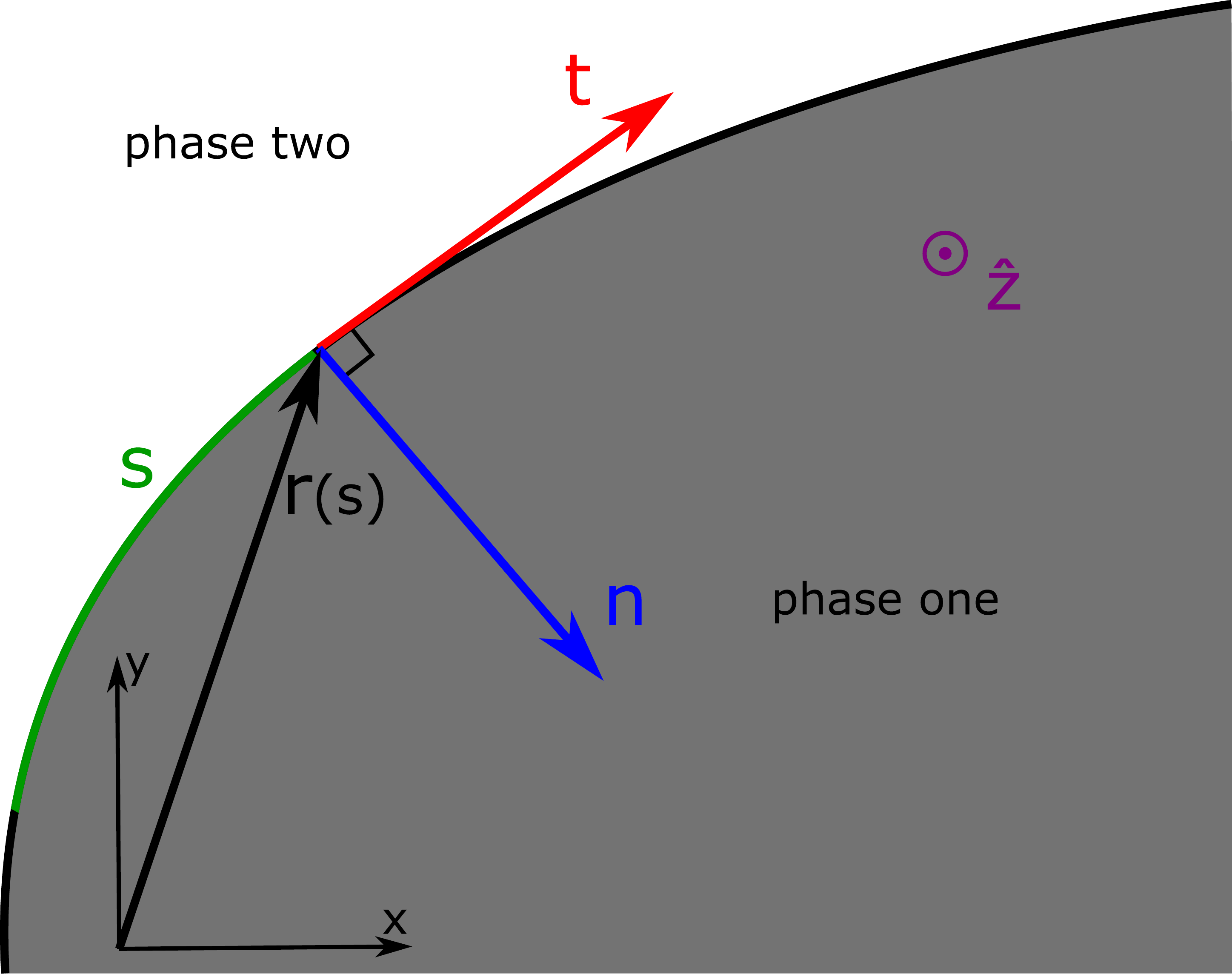}
\caption{
Parametrization of the interface curve between phases one (gray) and two (white).
Here $\textbf{r}$ is the coordinate of the point on the interface parameterized by arc length $s$.
$\textbf{t}$ is unitary vector tangent to the curve $\textbf{t} = \textbf{r}'$.
$\textbf{n}$ is unitary vector orthogonal to the curve $\textbf{n} = \textbf{t}' / \kappa = \textbf{t} \times \hat{\textbf{z}}$.
}
\label{fig_vectors_parametrization}
\end{figure}

Let us next consider a phase-separation-like process but in a different limit where instead $c$ is not small in general and changes very fast in space, i.e. the width of the interface between the phases is negligibly small.
Hence we can approximately set $c = \pm1$ everywhere.
Namely, now configuration is uniquely defined by coordinates of interfaces between  two phases labeled by $c = 1$ and $c = -1$.
There could be multiple disconnected interfaces.
Let us enumerate the interfaces by the index $i = 1, .. N$ and parameterize curves associated with the interfaces by arc length $s$, such that each interface is given by $\textbf{r}_i(s)$, see \figref{fig_vectors_parametrization}.
Hence the energy functional is given by:
\begin{equation}
F[c(\textbf{r})] = \sum_i G[\textbf{r}_i(s)],
\end{equation}

where $G$ is a new energy functional that depends on the shape and size of the given interface.
Now, let us phenomenologically derive the explicit form of the energy functional $G$ in the spirit of the Cahn-Hilliard model \eqref{CH}.
To do that we need to ensure that the model satisfies different symmetry conditions.
Also, we need to determine what will play the role of the order parameter.
In analogy with \eqref{CH} we can write:
\begin{equation}\label{Grs}
G[\textbf{r}(s)] = \int g\left( \textbf{r}(s), \textbf{r}'(s), \textbf{r}''(s), .. \right) ds
\end{equation}

The model should be translationally invariant.
Hence $g$ shouldn't depend on $\textbf{r}$.
Namely, shifting the patch of one phase shouldn't change the energy.
Next $\textbf{r}' \equiv \textbf{t}$ where $\textbf{t}(s)$ is unitary vector tangent to the interface, see \figref{fig_vectors_parametrization}.
We demand that the model should be rotationally invariant. Hence $g$ should not  depend on $\textbf{t}$.
The next derivative gives $\textbf{r}'' = \textbf{t}' = \kappa \textbf{n}$, where $\kappa(s)$ is the signed curvature of the interface curve.
Vector normal to the curve is $\textbf{n} = \textbf{t} \times \hat{\textbf{z}}$, where $\hat{\textbf{z}} = (0, 0, 1)$  is unit vector orthogonal to $xy$ plane.
Since $\textbf{n}' = - \kappa \textbf{t}$ and $\textbf{t}' = \kappa \textbf{n}$ any higher order derivatives of $\textbf{r}$ will be spanned by vectors $\textbf{n}$ and $\textbf{t}$ and depend on curvature and it's derivatives.
Namely, $\textbf{r}^{(n)} = a \textbf{n} + b \textbf{t}$ where $a,\ b$ are functions of $\kappa, \kappa' ..$.
For example, $\textbf{r}'''  = \kappa' \textbf{n} - \kappa^2 \textbf{t}$.
This means that model \eqref{Grs} depends only on signed curvature $\kappa$ and its derivatives.
Hence it's the natural equivalent of the order parameter in such model\footnote{Consider another way to see how curvature appears in this model.
When expanding Cahn-Hilliard model to higher orders in derivatives one obtains that, for example, for $c(r)$ we get $\nabla^2 c = c'' + \kappa c'$. After integrating orthogonal to the interface one obtains a model that is functional of curvature. See \cite{barkman2020ring} for a similar approximation.}.
Hence we can rewrite \eqref{Grs} in terms of $\kappa$:
\begin{equation}\label{Gks}
G[\kappa(s)] = \int g\left( \kappa(s), \kappa'(s), \kappa''(s), .. \right) ds
\end{equation}

\section{Small curvature $\kappa$ expansion}
Next, consider the case where $\kappa(s)$ is small and slowly changing function of $s$.
This leads to the expansion:
\begin{equation}\label{gk}
g = V(\kappa) + \gamma (\kappa')^2 + ..
\end{equation}

Let us consider the case where $\gamma$ is very large and hence $\kappa' = 0$.
In that case, the curvature is constant along the interface, making it a circle.
Note, that sign of the curvature depends on the phase.
Namely, if we set the disc of phase one (two) on phase two (one) background then we have positive (negative) curvature.
Let us consider different ground states that this model can have depending on the potential $V(\kappa)$.

(i) If $V > 0$ then the system will have one uniform phase.

(ii) If $V < 0$ then the system will make infinitely many interfaces.
Since we have interfaces of zero thickness (unlike the Cahn-Hilliard model) these interfaces will be infinitely close to each other and energy will diverge.

(iii) Energy of a single interface circle as a function of $\kappa$ changes sign and has a negative minimum.
Circle energy is then $U(\kappa) = \frac{2 \pi}{|\kappa|}V(\kappa)$.
Hence if $U(\kappa)$ has minimum with $U_{min} < 0$ and $U(0),\ U(+\infty) > 0$ the ground state of this model can represent a nontrivial configuration of interfaces.

(iv) Potential $V(\kappa)$ cannot be an even function of $\kappa$ to have a convergent minimum.
Since then there would be two minimums of equal energies for $\pm\kappa_{min}$.
Hence it would be beneficial to put a phase two disc with curvature $\kappa_2<0$ inside the phase one disc with $\kappa_1>0$.
Since this model has zero thickness interfaces these circles could have infinitely close curvatures $\kappa_1 \to - \kappa_2$.
This process can be repeated and infinitely many circles then would be inserted.
This means that energy would diverge.

\begin{figure}
\centering
\includegraphics[width=0.99\linewidth]{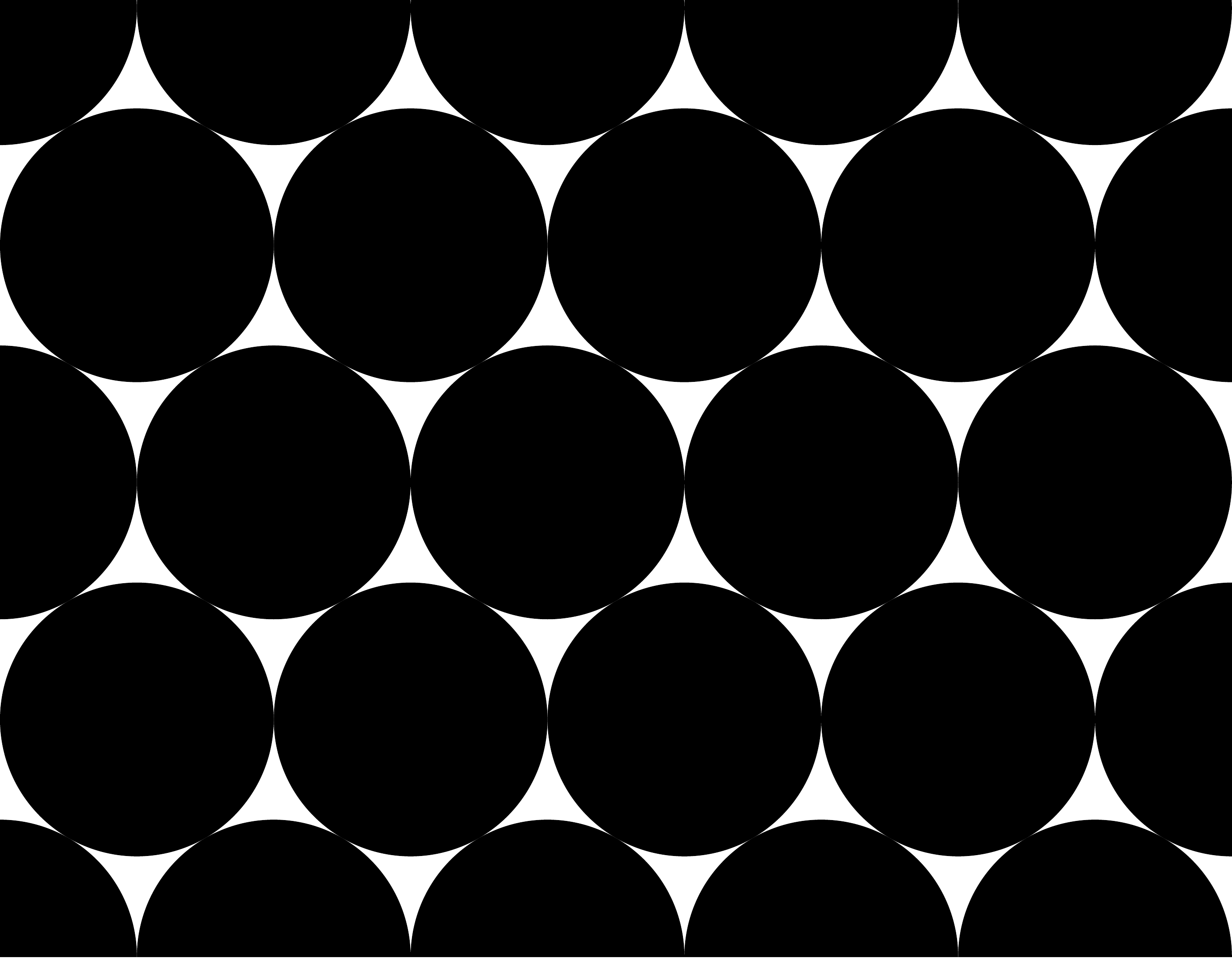}
\caption{
The crystalline ground state of the model \eqref{gk} and \eqref{Vk} in the form of hexagonal packing of discs.
Phase one (two) is colored black (white).
}
\label{fig_hexagonal_packing}
\end{figure}

Consider the simplest example 
\begin{equation}\label{Vk}
V(\kappa) = V_0 + V_1 \kappa + V_2 \kappa^2
\end{equation}
Hence $U(\kappa) = 2 \pi \left( \frac{V_0}{|\kappa|} + V_1 \text{sign}(\kappa) + V_2 |\kappa| \right)$, where $V_0,\ V_2 > 0$ and $V_1 \neq 0$.
We can always rescale the model to get rid of $V_0$ and $V_2$.
The sign of $V_1$ just sets whether positive or negative $\kappa$ will be preferred.
Let us set $V_1 = - v$, where $v >0$ resulting in circular interface energy:
\begin{equation}\label{Uk}
U(\kappa) = 2 \pi \left( \frac{1}{|\kappa|} - \text{sign}(\kappa) v + |\kappa| \right)
\end{equation}
This energy is minimal for $\kappa_{min} = 1$ and equals $U_{min} = 2 \pi (2 - v)$.
For $v > 2$ it can have hexagonal lattice of circles as ground state with $\kappa_{hex} = \left( v - \sqrt{v^2 - 3} \right)^{-1}$, see \figref{fig_hexagonal_packing}.
Where $\kappa_{hex} < \kappa_{min}$ since $\kappa_{hex}$ is found to minimize energy density $\rho = 3 U(\kappa) / S_{hex}$, with hexagon area $S_{hex} = 6 \sqrt{3} / \kappa^2$.
See \appref{app_kProof} for a comparison of energy of this state to other packings, where we prove that it has lower energy than all other compact packings of discs and all packings with lower density.
Compact packing is a packing where every pair of discs in contact is in mutual contact with two other discs. 

\section{Expansion in small curvature radius $R$ }
Let us consider the case opposite to the one studied in the previous section.
Namely, here we assume that curvature $\kappa$ is rather large.
In this case we can expand in signed curvature radius $R \equiv 1 / \kappa$.
We obtain expansion similar to \eqref{gk}:
\begin{equation}\label{gR}
g = V(R) + \gamma (R')^2 + ..
\end{equation}
Term proportional to $(R')^2$ here plays similar role as $(\kappa')^2$ in \eqref{gk}.
The only difference is that $(R')^2$ diverges if the interface is not convex -- meaning that $R$ along the interface should be sign definite.
Otherwise, it plays the same role of fixing the shape of the interface.
In this section, we also assume that $\gamma$ is rather large so that interfaces form circles.

\begin{figure}
\centering
\includegraphics[width=0.99\linewidth]{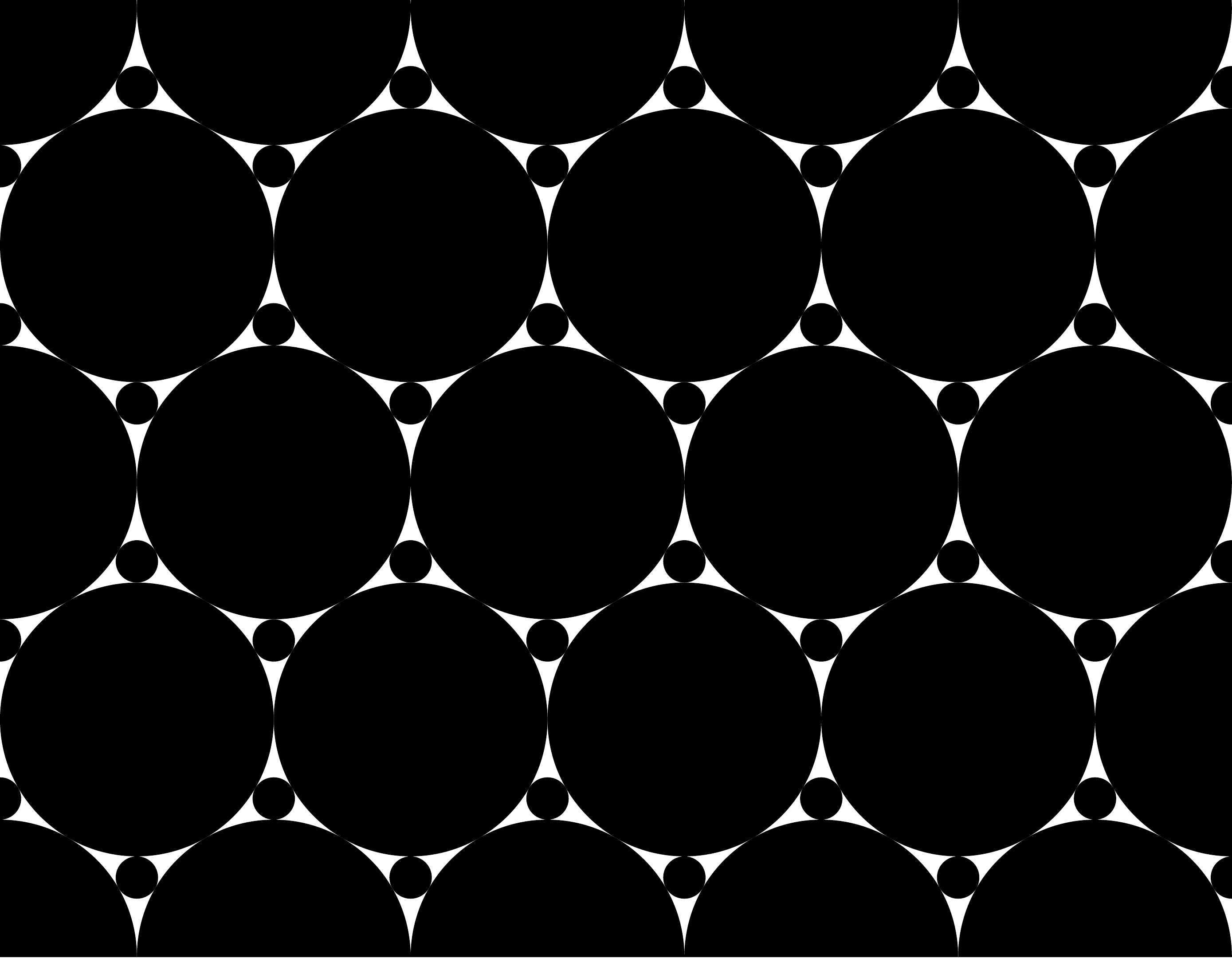}
\caption{
The crystalline ground state of the model \eqref{gR} and \eqref{VR} in the form of hexagonal packing of discs with smaller discs in between the large discs.
Phase one (two) is colored black (white).
}
\label{fig_hexagonal_packing_1}
\end{figure}

First, let us study the simplest (rescaled) model:
\begin{equation}\label{VR}
V(R) = 1 - v R + R^2
\end{equation}
which leads to single circle energy $U(R) = 2 \pi |R| \left( 1 - v R + R^2 \right)$.
We suppose that this model has various ground states.
For $0 < v < 2$ ground state is uniform single phase.
At $v > 2$ the model spontaneously breaks transitional symmetry: for $2 < v < 6.20..$ it is a hexagonal lattice \figref{fig_hexagonal_packing} with disc radius $R = 1.14..$.
Another phase transition happens at $v = 6.20$.
For $6.20.. < v < 13.17..$ it is a hexagonal lattice with an additional set of smaller discs \figref{fig_hexagonal_packing_1}.
Larger discs have radius $R = 1.29..$.
To check that we compared the energy densities of different packings of the discs \cite{kennedy2006compact}, see \appref{app_Rcomp}.
This pattern may continue by adding more and more smaller circles.

\begin{figure}
\centering
\includegraphics[width=0.99\linewidth]{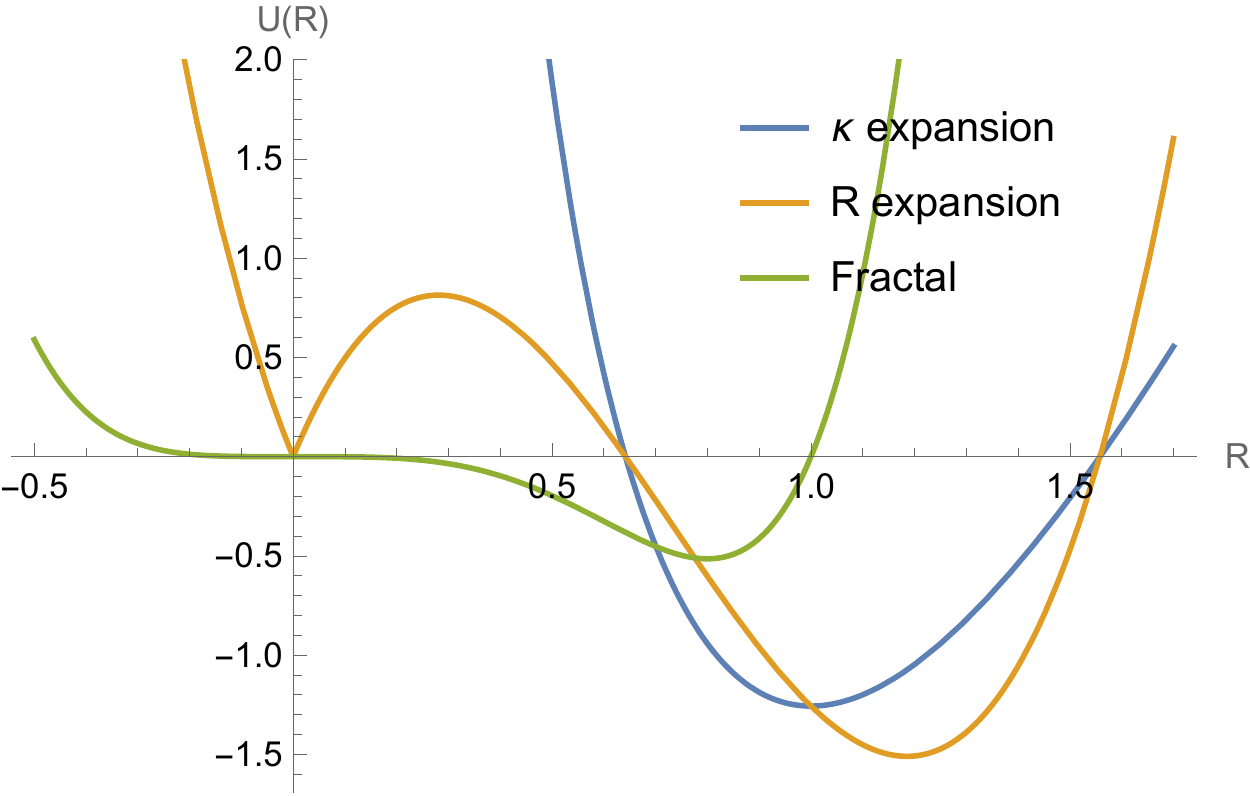}
\caption{
The energy of a disc $U(R)$ as a function of its signed curvature radius $R$ for different models.
Namely, the models, which are obtained by expanding in powers of curvature $\kappa$ \eqref{Uk} (blue) and radius $R$ \eqref{VR} (orange).
Here we fixed $v = 2.2$ and hence they have hexagonal packing of discs as the ground state.
Fractal crystal is obtained as the ground state in model \eqref{VR_frac} (green).
}
\label{fig_Us}
\end{figure}

\section{The ground state fractal crystal}
In this section, we demonstrate a model where the translation invariance breaks down to the ground state fractal crystal.
To achieve that it has to be energetically beneficial to add circles to any-sized gaps between already placed circles.
It means that we can set $V(R) \to 0^-$ for $R \to 0^+$.
Hence let us assume that potential $V(R) \to - R^\alpha$ in that limit.
Energy density for hexagonal lattice of small circles is then $\rho = 2 \pi |R| V(R) / S_{hex} \to - R^{\alpha - 1}$ for $R \to 0^+$.
So we obtain the condition that $\alpha \geq 1$.
Otherwise, the energy diverges as many circles of size $R \to 0^+$ populate the system.

Note, that $\alpha = 1$ is a special case since in $R \to 0^+$ limit all packings fully covering the plane have the same energy.
Hence if the subleading order in energy density is positive the ground state energy will diverge by the inclusion of infinitely small discs.
If the subleading order is negative ground state can be realized by some nontrivial fractal packing of discs.

One of the simplest options is to set $\alpha = 3$ and the potential to be:
\begin{equation}\label{VR_frac}
V(R) = - R^3 + R^4
\end{equation}
This model has circle energy given by $U(R) = 2 \pi |R| \left( - R^3 + R^4 \right)$.
For comparison of $U(R)$ plots in different models, studied in this work, see \figref{fig_Us}.

Firstly, it is easy to see that discs occupying the plane in model \eqref{VR_frac} will fully cover the plane.
This can be seen as follows: if they do not and there are some gaps between discs -- the energy can be decreased by placing smaller discs there.

Next, we can show which type of packing will be present for discs in the limit $R \to 0$.
To that end, consider an empty gap of area $S \to 0$ between already placed discs.
We want to find what type of disc packing will give the lowest energy for this gap.
Energy in this limit in general is 
\begin{equation}
E = - A_k \ \ \ {\rm  with} \ \ \  k > 2
\end{equation}
For the case \eqref{VR_frac} we have $k = 4$.
$A_k$ is a sum of radii to power $k$, which is given by \cite{gilbert1964randomly}:
\begin{equation}\label{Ak}
A_k = \sum_{i = 0}^{+\infty} n_i R_i^k = - c_k \int_0^{R_0} R^k n'(R) dR
\end{equation}
where $c_k$ are constants, $n_i$ is number of discs with radius $R_i$, sorted such that $R_0 > R_1 > ..$.
Whereas $n(R)$ is number of discs with radii $r$ such that $R_0 \geq r \geq R$.
For a given packing of circles, for $R \to 0$ it is possible to show \cite{melzak1969solid} that
\begin{equation}
n(R) = c R^{-d}
\end{equation}
where $d$ is the Hausdorff dimension of the packing  and $c$ is some other constant characterising it.
Hence $A_k$ can be estimated as
\begin{equation}\label{Ak_sol}
A_k = \frac{c_k c d}{k - d} R_0^{k - d}
\end{equation}
where parameters $c,\ d$ and $R_0$ depend on the packing.
Using relation $\pi A_2 = S$ we can eliminate the  parameter $c$:
\begin{equation}\label{Ak_asymp}
A_k = b \frac{2 - d}{k - d} R_0^{k - 2}
\end{equation}
where packing independent constant $b = S c_k / c_2$.
From \eqref{Ak_asymp} we see that maximum of $A_k$ and hence minimum of energy $E = - A_k$ is achieved for maximal $R_0$ and minimal $d$.
Maximal $R_0$ means that the largest disc should be as large as the gap allows (which corresponds to Apollonian packing).
While as was shown in \cite{melzak1969solid} $d \in [d_A, 2]$ for various disc packings, where $d_A \simeq 1.3056867..$ is the dimension of Apollonian packing.
Hence we see that for $R \to 0$ the ground state is fractal Apollonian packing of discs.

\begin{figure}
\centering
\includegraphics[width=0.99\linewidth]{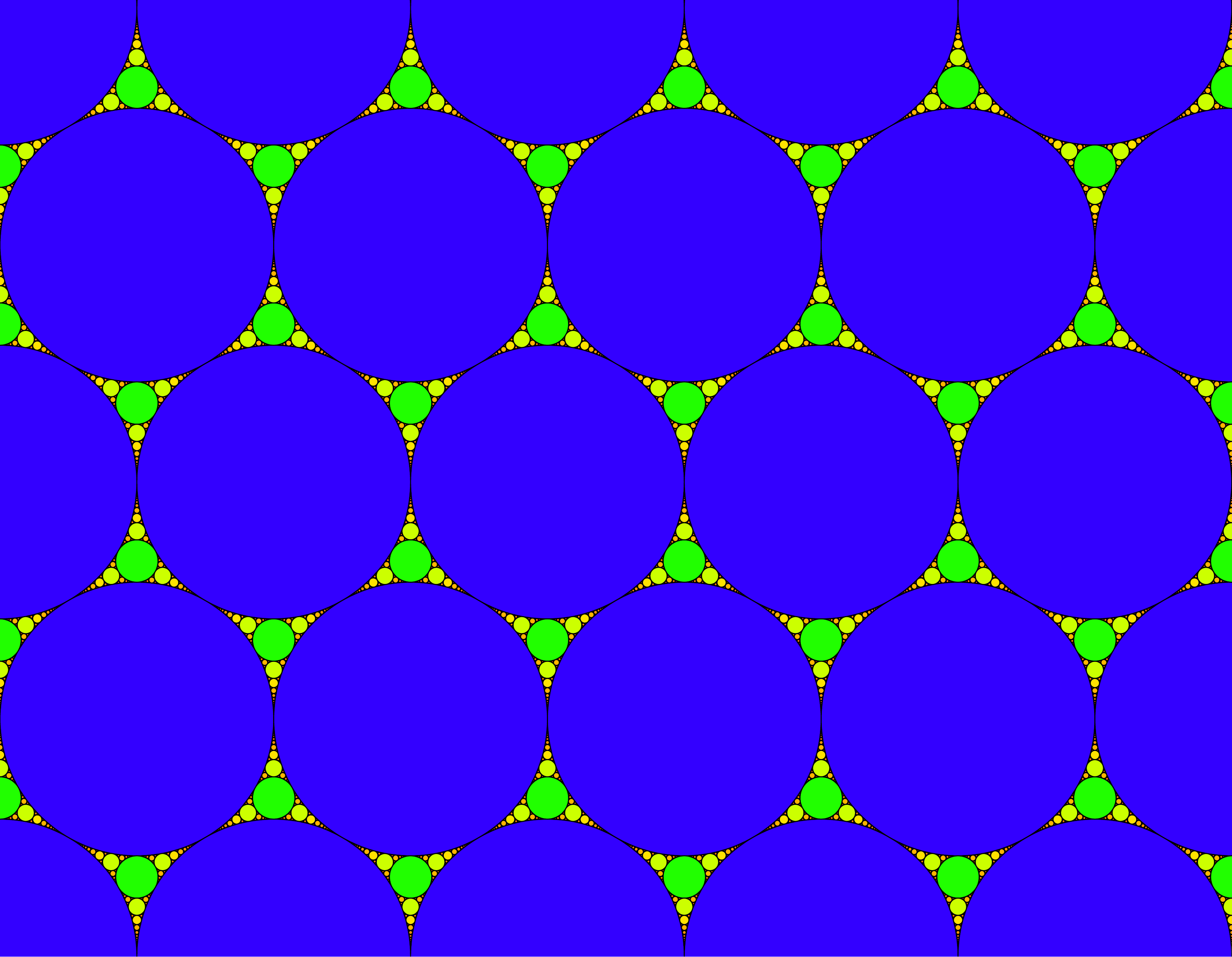}
\includegraphics[width=0.99\linewidth]{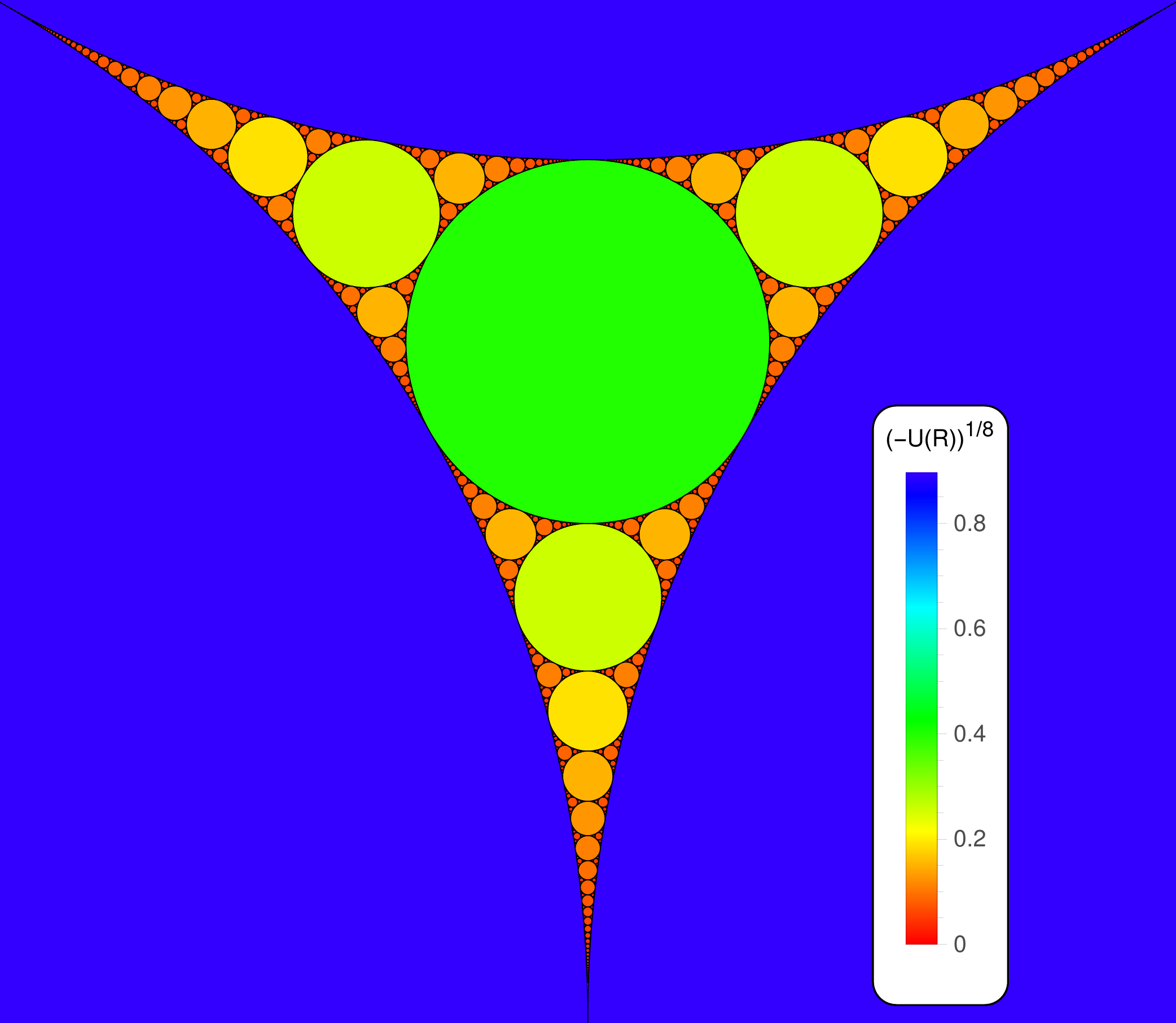}
\caption{
Fractal crystal in the form of a periodic Apollonian packing arising as the energy minimum state in the model \eqref{gR} and \eqref{VR_frac}.
Colors correspond to different energies of phase one discs.
For some finite number of generations of circles of a certain size, the second phase occupies the gaps between them.
As smaller circles are included the fraction of the second phase goes to zero.
The interfaces between the phases are drawn in black.
}
\label{fig_apollonian_packing}
\end{figure}

For the model \eqref{VR_frac} we propose a candidate for global minimum \figref{fig_apollonian_packing}.
Where radius of the biggest circle is $R = \left. \frac{2 A_4}{3 A_5} \right|_{R_0 = 1} = 0.667379..$ and energy density $\rho = -  \left. \frac{8 \pi A_4^3}{27 S A_5^2}  \right|_{R_0 = 1} = -0.269622..$.

\section{Phase transition  between fractal and uniform states}
To study transition between fractal and uniform phases consider the following modification of model \eqref{VR_frac}:
\begin{equation}\label{VR_frac_uni}
V(R) = a R^2 - R^3 + R^4
\end{equation}
Where for $a = 0$ fractal phase is the ground state.
As $a$ is increased up to $1 / 4$ supposedly larger and larger discs are removed from the fractal.
For $a > 1 / 4$ \eqref{VR_frac_uni} has uniform ground state with energy density $\rho = 0$.
We can define order parameter in this case as density of area which is not occupied by discs
\begin{equation}
\sigma = (S_{total} - S_{discs}) / S_{total}.
\end{equation}

Now let us introduce a  critical exponent $\omega$ defined by 
\begin{equation}
\sigma \propto a^\omega \ \  {\rm for} \ \ a \to 0.
\end{equation}
Consider configuration \figref{fig_apollonian_packing} with small discs of radius $R < R_n$ removed.
Similar to \eqref{Ak} and \eqref{Ak_sol} in the limit $a \to 0$ we can compute $\sigma$ as:
\begin{equation}
\sigma = \pi S_{total}^{-1} \sum_{i = n + 1}^{+\infty} n_i R_i^2 = \frac{c_2 c d_A}{2 - d_A} R_{n + 1}^{2 - d_A} \propto R_{n + 1}^{2 - d_A}
\end{equation}
Hence we need to find the relation between $R_{n + 1}$ and $a$ parameter.
To do so note, that, energy density is given by $\rho^n = 2 \pi \frac{a A_3^n R_0^3 - A_4^n R_0^4 + A_5^n R_0^5}{R_0^2 S_0}$.
Where $A_k^n = \sum_{i = 0}^n n_i \eta_i^k$ with $\eta_i = R_i / R_0$ and unit cell area $S_0$ are rescaled in terms of the radius of the largest disc $R_0$.
Then we can minimise $\rho_n$ in terms of $R_0$ and expanding in $a$ we get:
\begin{equation}
\rho_{min}^n = \frac{4 \pi A_4^n}{3 S_0 A_5^n} \left( -\frac{2 (A_4^n)^2}{9 A_5^n} + a A_3^n \right) + O(a^2)
\end{equation}
Hence when  $a$ is decreased configuration with $n + 1$ discs will become the ground state instead of the configuration with $n$ discs when $\rho_{min}^n = \rho_{min}^{n + 1}$.
Solving the later for $a$, which we denote $a_{n + 1}$ in this case:
\begin{equation}
a_{n + 1} = \frac{2 A_4^n}{3 A_5^n} \eta_{n + 1} + .. \propto R_{n + 1}
\end{equation}
which means that the critical exponent:
\begin{equation}
\sigma \propto a^\omega,\ \ \text{with}\ \ \omega = 2 - d_A \simeq 0.6943..
\end{equation}

\section{Effects of fluctuations}
The fact that fractal, in our model arises as a ground state, rises the interesting question about the effects of the fluctuations.

Here, we speculate how thermal fluctuations can induce a new kind of hierarchical phase transitions and how they can be characterized.
The discussion in this section is more general than the concrete realization of ground state fractal considered in the previous section, rather we want to discuss how to characterize the melting of a fractal.
Suppose that, for nonzero temperature, the order is destroyed for discs with a radius smaller than $R \propto T^{\frac{1}{\alpha + 1}}$, where $\alpha$ is the power of the energy potential \eqref{VR_frac}.
This is of course just an assumption, that may require a generalization of the model to realize.
If it or a similar situation does realize, then increasing temperature from zero will disorder bigger and bigger discs.

\begin{figure}
\centering
\includegraphics[width=0.99\linewidth]{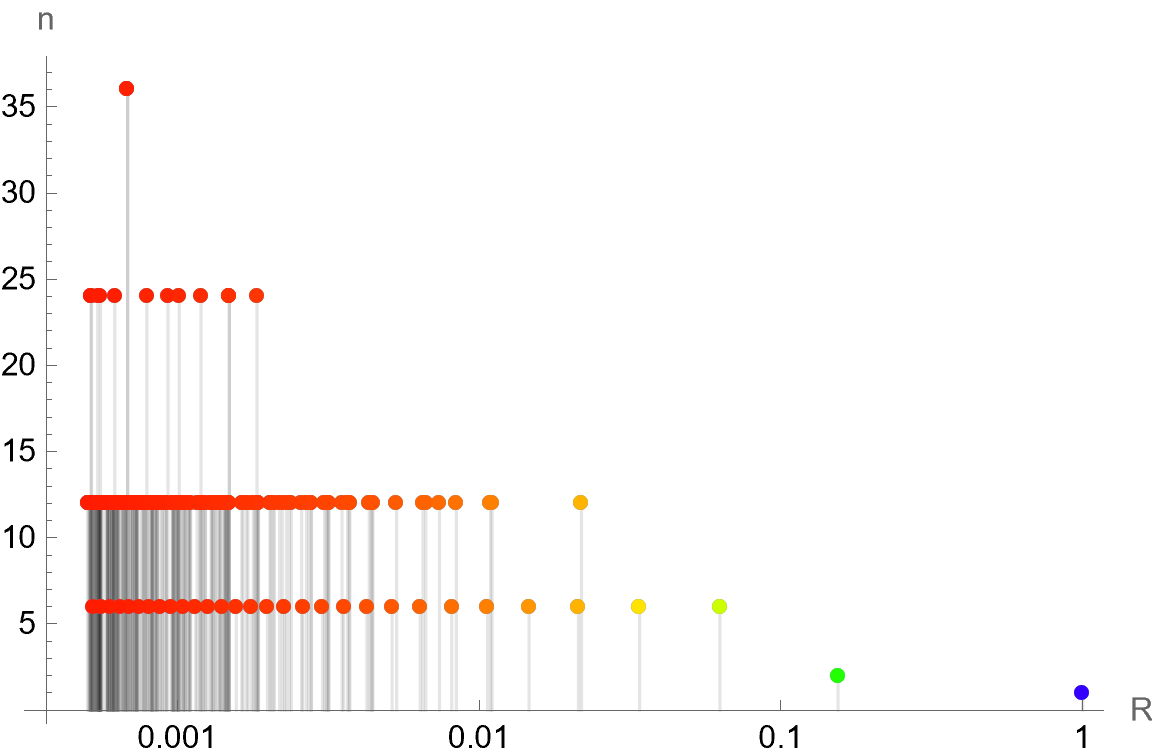}
\caption{
Number of discs $n$ of given radius $R$ in the unit cell of a fractal crystal \figref{fig_apollonian_packing}.
At finite temperatures, we expect peaks to be widened by thermal fluctuations.
Nonetheless, by searching for peaks in $n(R)$ distribution, one can identify radii that can be used to define corresponding order parameters in terms of structure factors \figref{fig_structure_factors}.
}
\label{fig_n_R}
\end{figure}

To describe such a transition, we need a new kind of order parameter with an infinite number of components that describes a ground state fractal and fluctuations therein.
We propose the following order parameters.
Consider the case where fluctuations affect both the positions and radii of the discs. 
To characterize the ordering, one can plot the number of discs $n$ in the system as a function of their radius $R$.
The corresponding plot for a zero-temperature case is given in \figref{fig_n_R}.
Then this plot can be used to identify peaks corresponding to radii of discs that may form some crystal structure or be in a liquid state.
Then, even if there are size fluctuations, the peaks allow grouping of the discs into various ``generations".
Next, by plotting the structure factor for discs in the chosen peak/generation, one can see whether they are disordered (i.e. liquid or glass state) or form a crystal.
We define this structure factor for a given peak as follows:
\begin{equation}\label{structure_factor}
S(\textbf{k}) = \left| \int_{\text{phase one}} e^{\ii \textbf{k} \cdot \textbf{r}} d\textbf{r} \right| = 2 \pi \left| \sum_i \frac{J_1(k R_i)}{k R_i} e^{\ii \textbf{k} \cdot \textbf{r}_i} \right|
\end{equation}
where, without loss of generality, we set phase one (two) to be $1$ ($0$), so the integral is only over phase one -- namely, over discs in the given peak.
The integral is solved by Bessel functions of the first kind $J_1$.
So overall structure factor depends on the radii of discs $R_i$ and the positions of their centers $r_i$.

We compute structure factor \eqref{structure_factor} for zero temperature fractal crystal state \figref{fig_apollonian_packing} for given discs size $R_i$:
\begin{equation}\label{sf_0}
S(k) = 2 \pi \frac{J_1(k R_i)}{k R_i} \left| \sum_j e^{\ii \textbf{k} \cdot \textbf{r}_j^0} \right| \delta_{2 \pi}(\textbf{k} \cdot \textbf{a}_1) \delta_{2 \pi}(\textbf{k} \cdot \textbf{a}_2)
\end{equation}
where delta functions are up to $2 \pi$ addition: $\delta_{2 \pi}(x) = \sum_{n = -\infty}^{+\infty} \delta(x + 2 \pi n)$ -- which gives delta functions at sites in the reciprocal lattice.
Primitive vectors of direct hexagonal lattice are $\textbf{a}_1 = R_0 (2, 0)$ and $\textbf{a}_2 = R_0 (1, \sqrt{3})$.
While $\textbf{r}_j^0$ are positions of disc centers in the unit cell, which are indexed by $j$.

\begin{figure}
\centering
\includegraphics[width=0.49\linewidth]{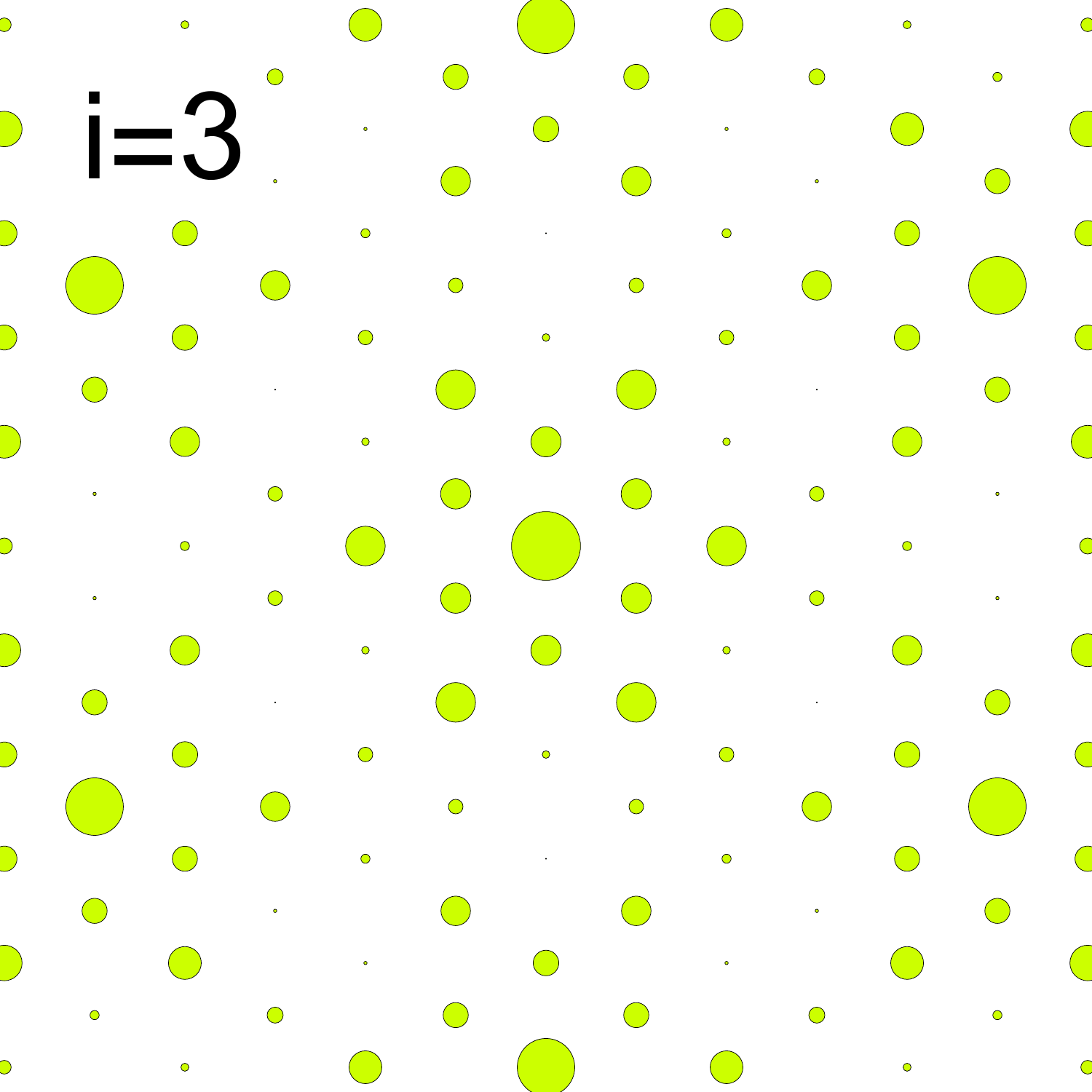}
\includegraphics[width=0.49\linewidth]{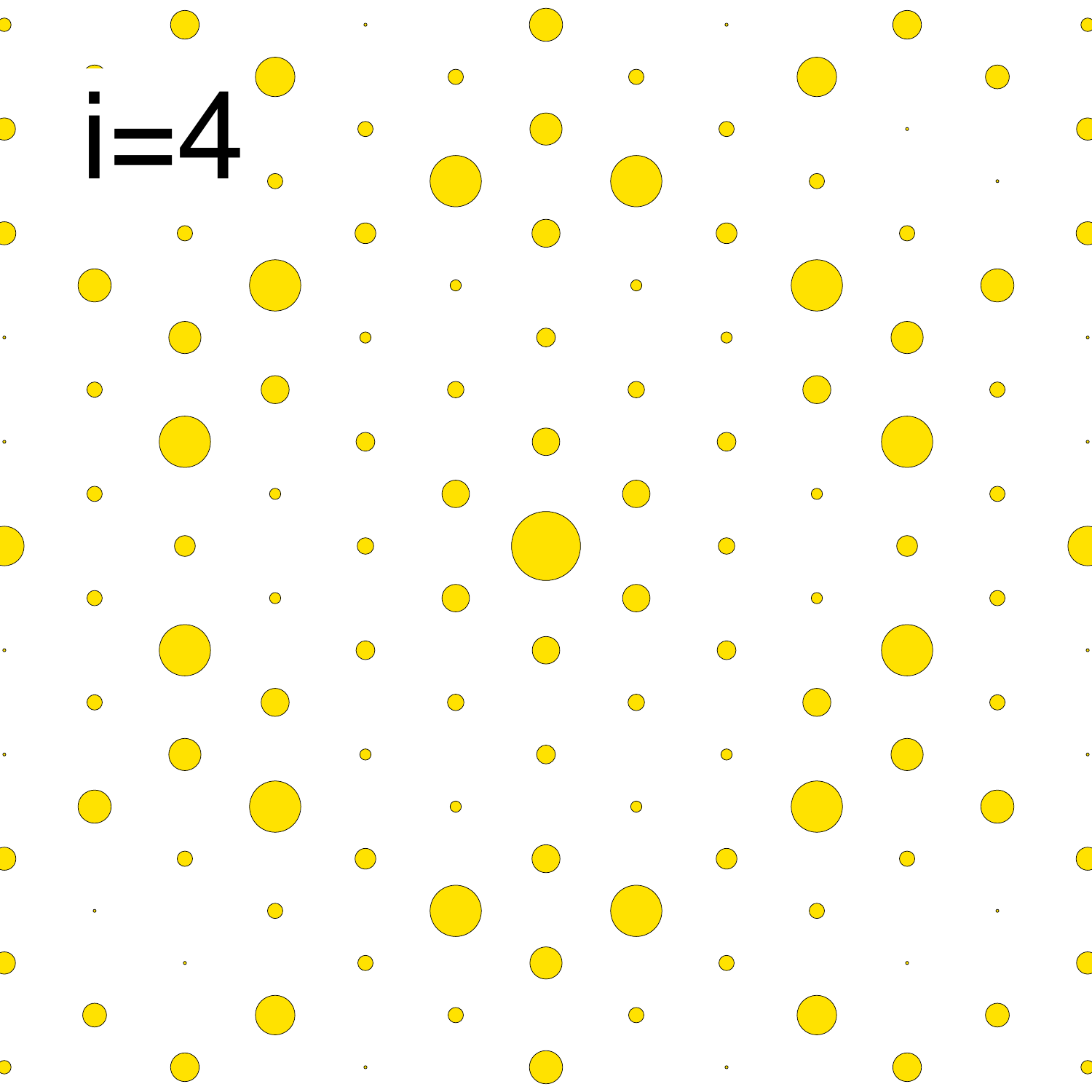}
\includegraphics[width=0.49\linewidth]{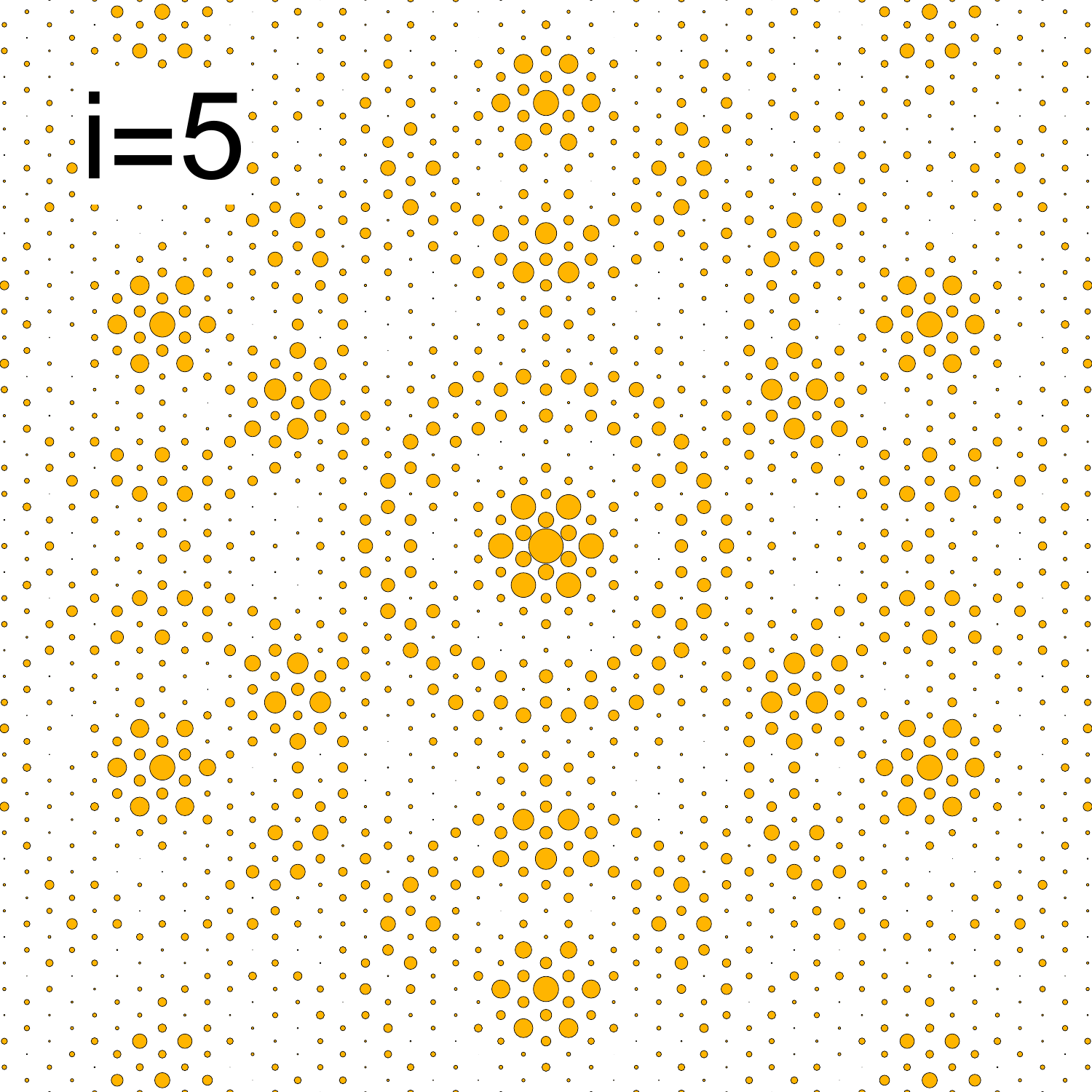}
\includegraphics[width=0.49\linewidth]{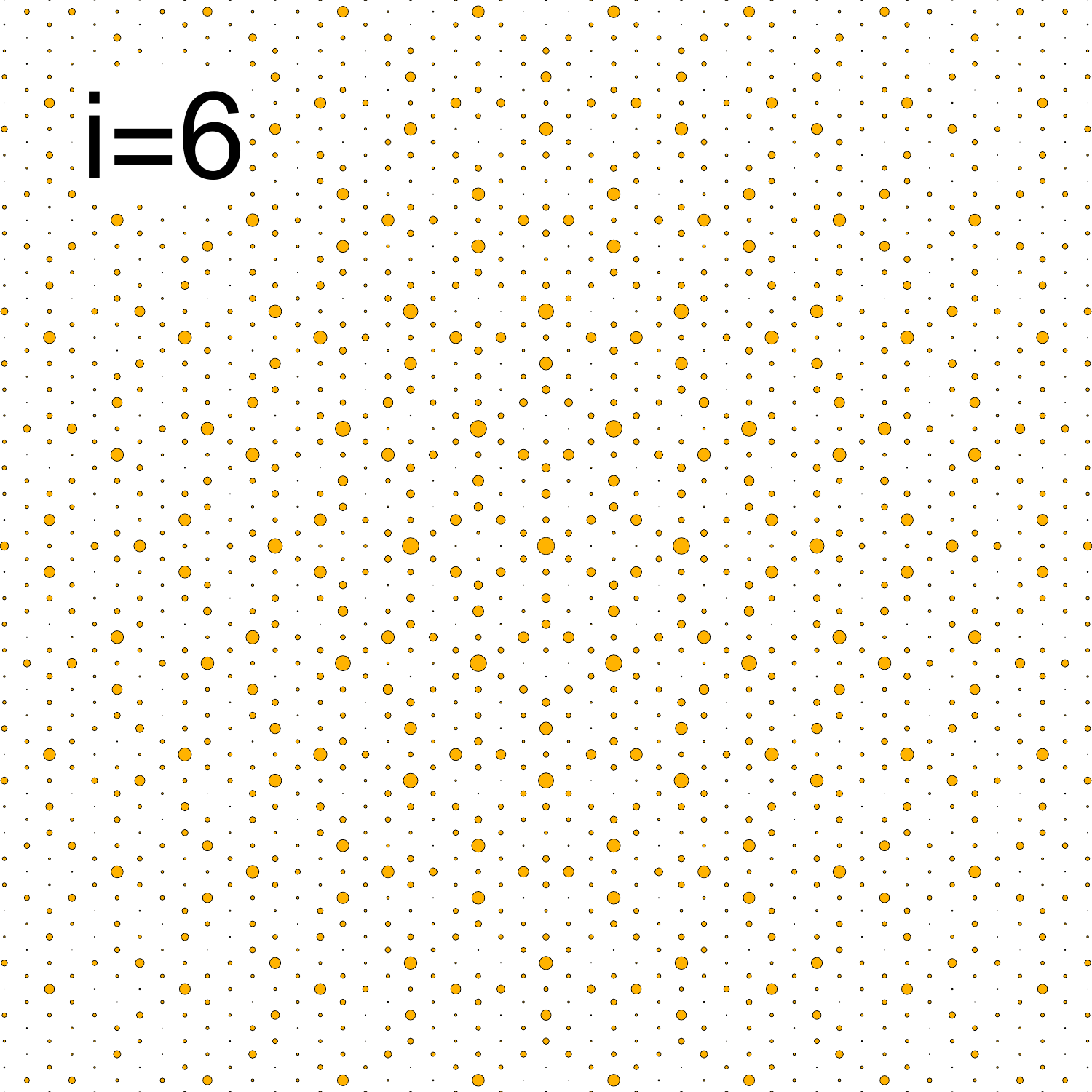}
\includegraphics[width=0.49\linewidth]{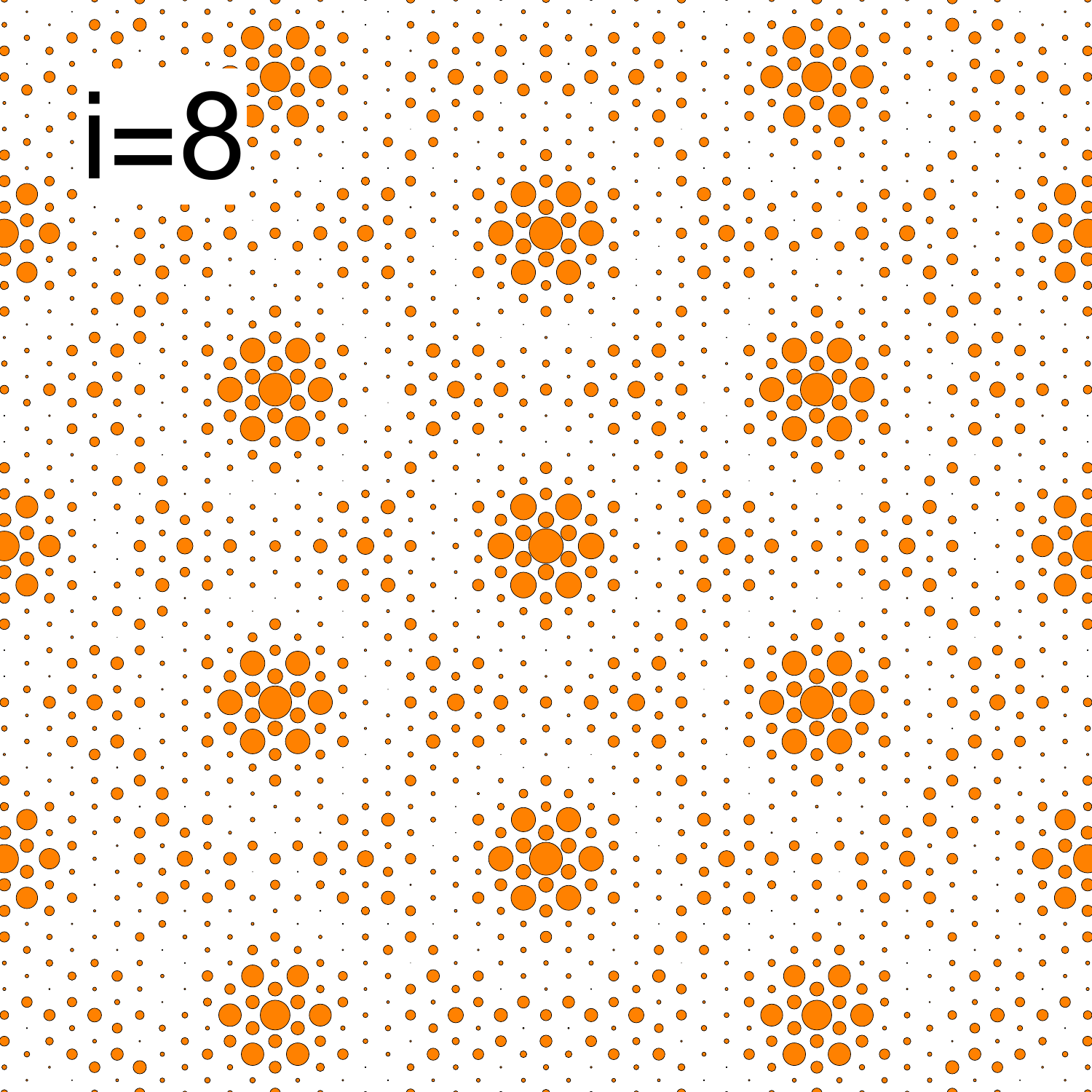}
\includegraphics[width=0.49\linewidth]{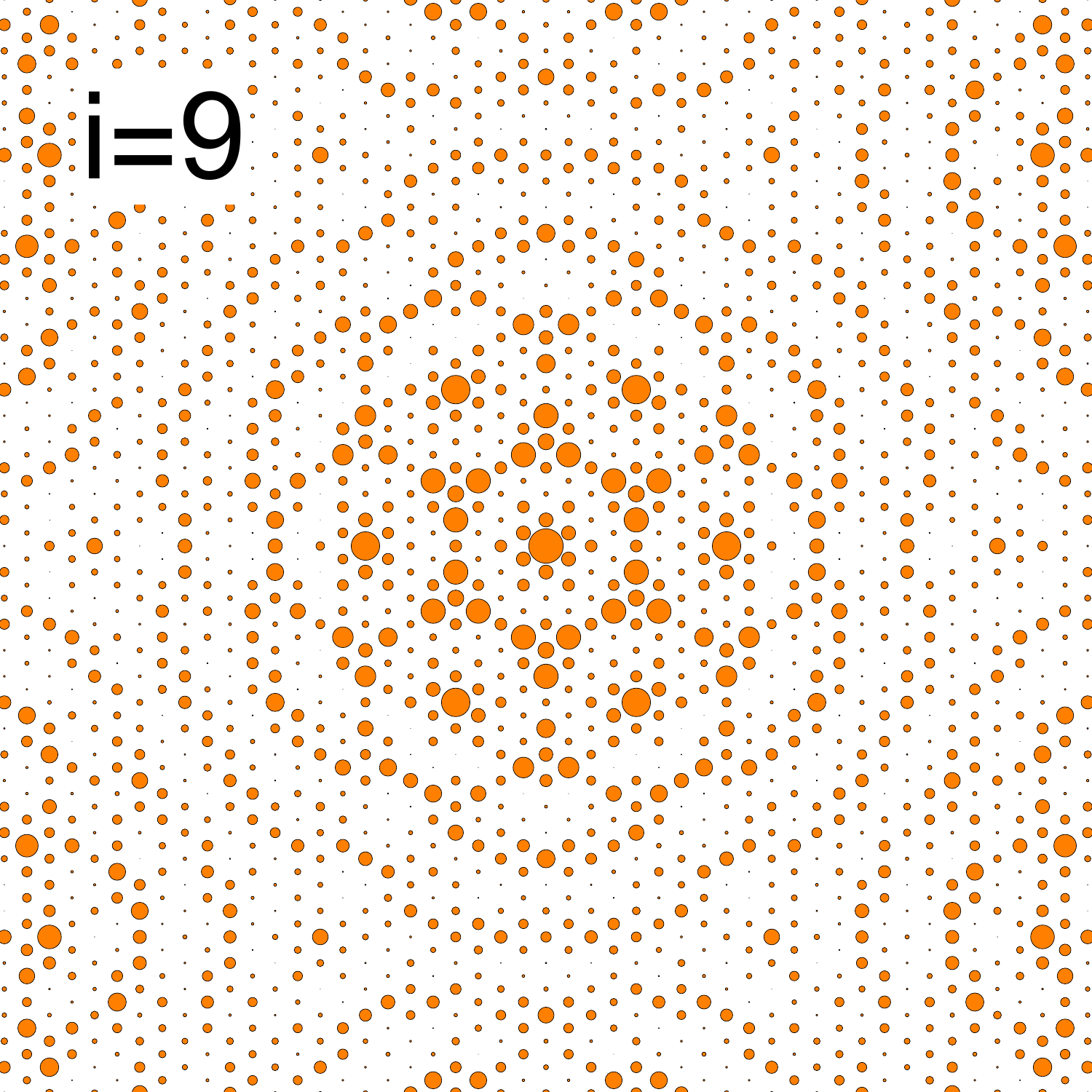}
\includegraphics[width=0.49\linewidth]{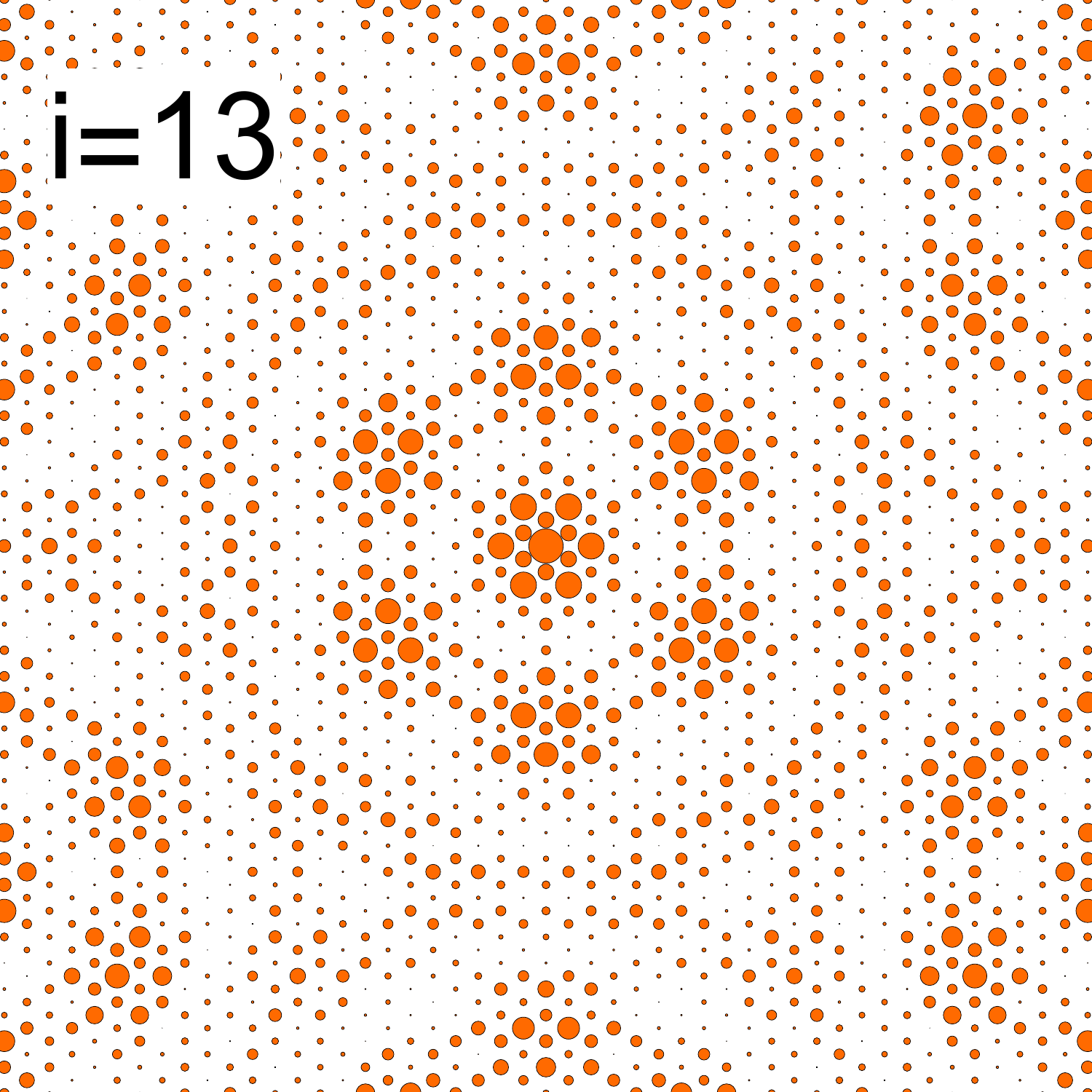}
\includegraphics[width=0.49\linewidth]{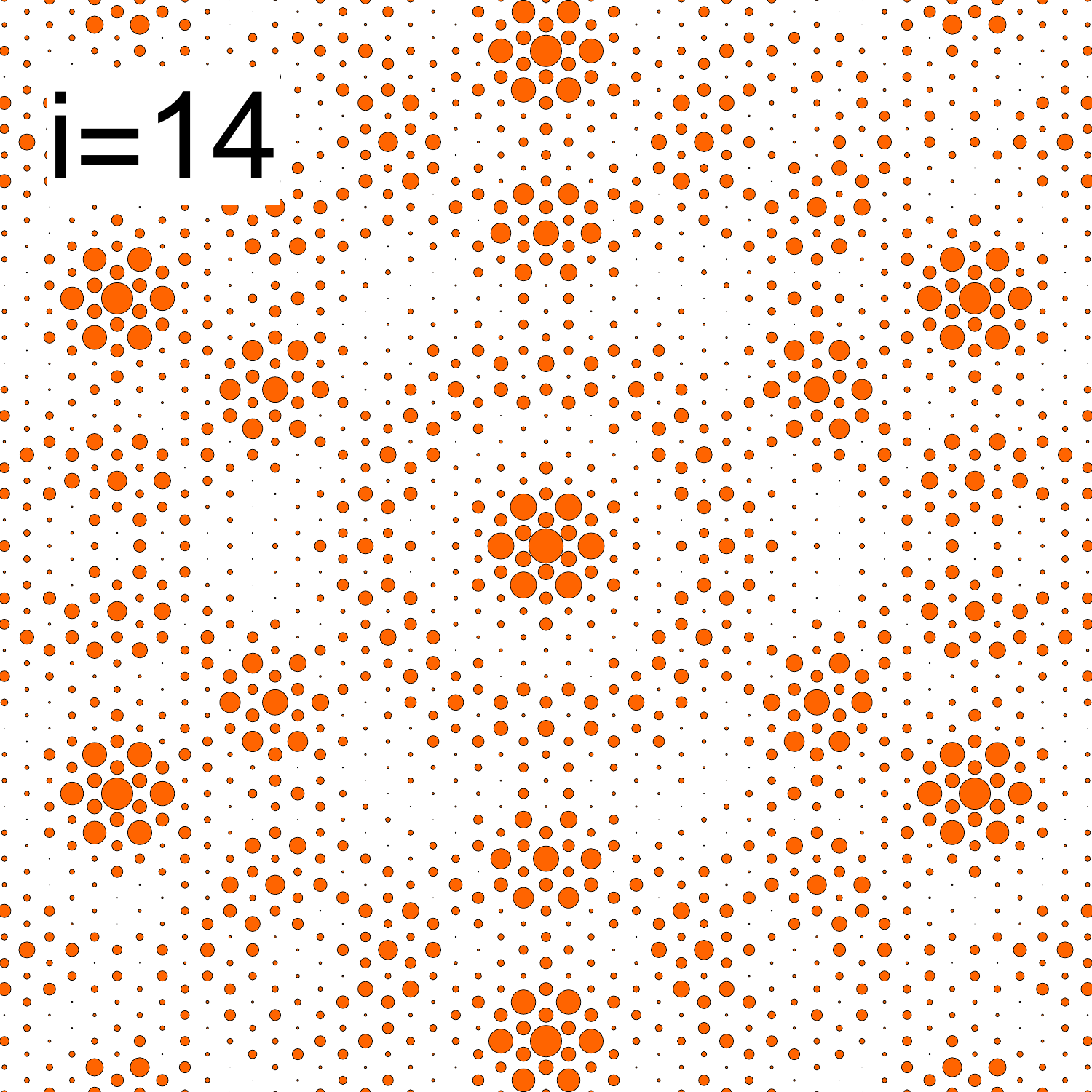}
\caption{
Examples of structure factors \eqref{sf_0} for discs of a given radius $R_i$.
Index $i$ enumerates discs from biggest to smallest.
Another way to put it is that $i$ enumerates peaks/generations of discs, and hence coloring corresponds to \figref{fig_apollonian_packing} and \figref{fig_n_R}.
Note, that in all cases, structure factor configuration is given by the hexagonal lattice of delta functions.
To illustrate that, we plot discs placed at the delta function positions, with the radii of the discs corresponding to the magnitudes of the delta function prefactors.
$i=3,4$ plots are zoomed in.
}
\label{fig_structure_factors}
\end{figure}

We plot these structure factors in \figref{fig_structure_factors}.
Note, that they form different patterns, which can help distinguish order parameters for different $i$.
However, some of them will look very similar, so peaks in $n(R)$ are needed to identify order parameters.
Hierarchical melting then would imply a sequence of losing the peaks of the structure factors of various colors, corresponding to different generations of the discs.

\section{Conclusions}
We presented the concept of the ground state fractal crystal, as a state generalizing crystalline order.
We phenomenologically derived the model that is defined on a two-dimensional continuous space and has a two-valued discrete field.
The model respects space translation symmetry.
The energy of this model is expressed as an integral over interfaces between two phases of function that depends on signed curvature $\kappa$ and its derivatives.
We demonstrated that the energy minimization in this model leads to the spontaneous breakdown of translation symmetry in the form of a crystal where each unit cell can be a fractal.

The model can be generalized to the situation with some finite interface thickness $d$.
Then the fractal structure will be present only down to order $d$.
This is similar to other fractals in physical systems which generically feature some microscopic cutoff length scale.

The open question is whether these states are realized in physical systems with complex order.
Such as the generalizations of the phases occurring in quantum Hall systems between stripe and bubble phases \cite{fogler2002stripe}, phases between a two-dimensional electron liquid and Wigner crystal \cite{spivak2004phases}, soft matter states or hierarchical structure formation in polydisperse vortex clusters \cite{meng2016phase}.

The fractal energy minimizers in a classical field theory that we find can in principle be related to quantum problems.
In our case, the energy minimizer is a crystal of Apollonian packing in each unit cell.
On the other hand, the fractals similar to integral Apollonian packing are related to Hofstadter Butterfly \cite{satija2016tale}, i.e. the energy spectrum of electrons in a magnetic field.

Finally, we note that we presented the simplest continuous space discrete field model that can be easily generalized.

For example, one can consider a three-dimensional version with principle curvature for two-dimensional interfaces between phases.
Next, more phases can be included, which will result in new types of interfaces.
Namely, for three or more phases one can also have a vertex in two dimensions (or vertices and lines in three dimensions) where three or more phases meet.
Energies of these lower-dimensional interfaces can be set in addition to functional that depends on curvature.
Moreover in order to obtain fractal packing of discs one in principle can have other models that somehow favor specific shape and bigger sizes of areas of a given phase, see \appref{app_Other2dFractals}.

For a one-dimensional system, the interface is just a point that has no curvature.
However, it is still possible to obtain some fractal patterns as the ground state in it, see \appref{app_1dFractals}.

Having a theory where a ground state is a fractal raises an interesting question about the nature of thermal fluctuations and melting.
It appears that thermal fluctuations in such systems might result in a hierarchical melting, with weak thermal fluctuations destroying the order of smaller-scale sublattices, with the system getting more disordered at increased temperatures. We have shown how one can construct an order parameter for such a hierarchical melting.

\begin{acknowledgments}
The work was supported by the Swedish Research Council Grants  2016-06122, 2018-03659.
This work was inspired by the video "Newton's Fractal (which Newton knew nothing about)" by 3Blue1Brown (\href{https://www.youtube.com/watch?v=-RdOwhmqP5s&ab_channel=3Blue1Brown}{link}).
We thank Mats Barkman, Sahal Kaushik, and Boris Svistunov for useful discussions.
\end{acknowledgments}

%

\clearpage
\appendix
\section{Comparison of different configurations in the $\kappa$ expansion model}\label{app_kProof}
Here we present analysis suggesting that the ground state of the model \eqref{gk}, \eqref{Uk} for $v \geq 2$ is  hexagonal packing of equal sized discs, see \figref{fig_hexagonal_packing}.

\subsection{Proof that hexagonal packing has lower energy than any compact packing of discs}
We rewrite energy of a disc \eqref{Uk} in terms of curvature radius $R \equiv 1/ \kappa$:
\begin{equation}\label{UkR}
U(R) = 2 \pi \left( R - v + \frac{1}{R} \right)
\end{equation}
where $R \geq 0$.
Consider a system of area $S \to +\infty$ (such that boundary effect is negligible) which has $n_i$ discs of radius $R_i$.
Index $i \in [0, N - 1]$ and $R_0 > R_1 > R_2 .. > R_{N - 1}$.
Then energy density that we want to minimize is:
\begin{equation}\label{rho_eqs}
\begin{gathered}
\rho = 2 \pi \frac{R_0 A_1 - v A_0 + \frac{A_{-1}}{R_0}}{R_0^2 S_0},\ \ \text{with}\ \ A_k = \sum_{i = 0}^{N - 1} n_i \eta_i^k \\
S = R_0^2 S_0,\ \ \eta_i = R_i / R_0
\end{gathered}
\end{equation}
where we wrote all lengths rescaled in terms of radius of the biggest disc $R_0$.
Now for given $\eta_i$ and $n_i$ we can minimize energy density with respect to $R_0$.
It gives:
\begin{equation}
\begin{gathered}
R_0 = \frac{A_0}{A_1} \left( v - \sqrt{v^2 - 3 \lambda^2} \right)\\
\rho = \alpha \lambda^3 f(v / \lambda)
\end{gathered}
\end{equation}
where 
\begin{equation}\label{alpha_lambda_f}
\begin{gathered}
\lambda = \sqrt{\frac{A_{-1} A_1}{A_0^2}},\ \ \alpha = \frac{A_0^3}{S_0 A_{-1}^2}\\
f(q) = 2 \pi \frac{q^2 - 2 - q \sqrt{q^2 - 3}}{\left( q - \sqrt{q^2 - 3} \right)^3}
\end{gathered}
\end{equation}

\begin{figure}
\centering
\includegraphics[width=0.99\linewidth]{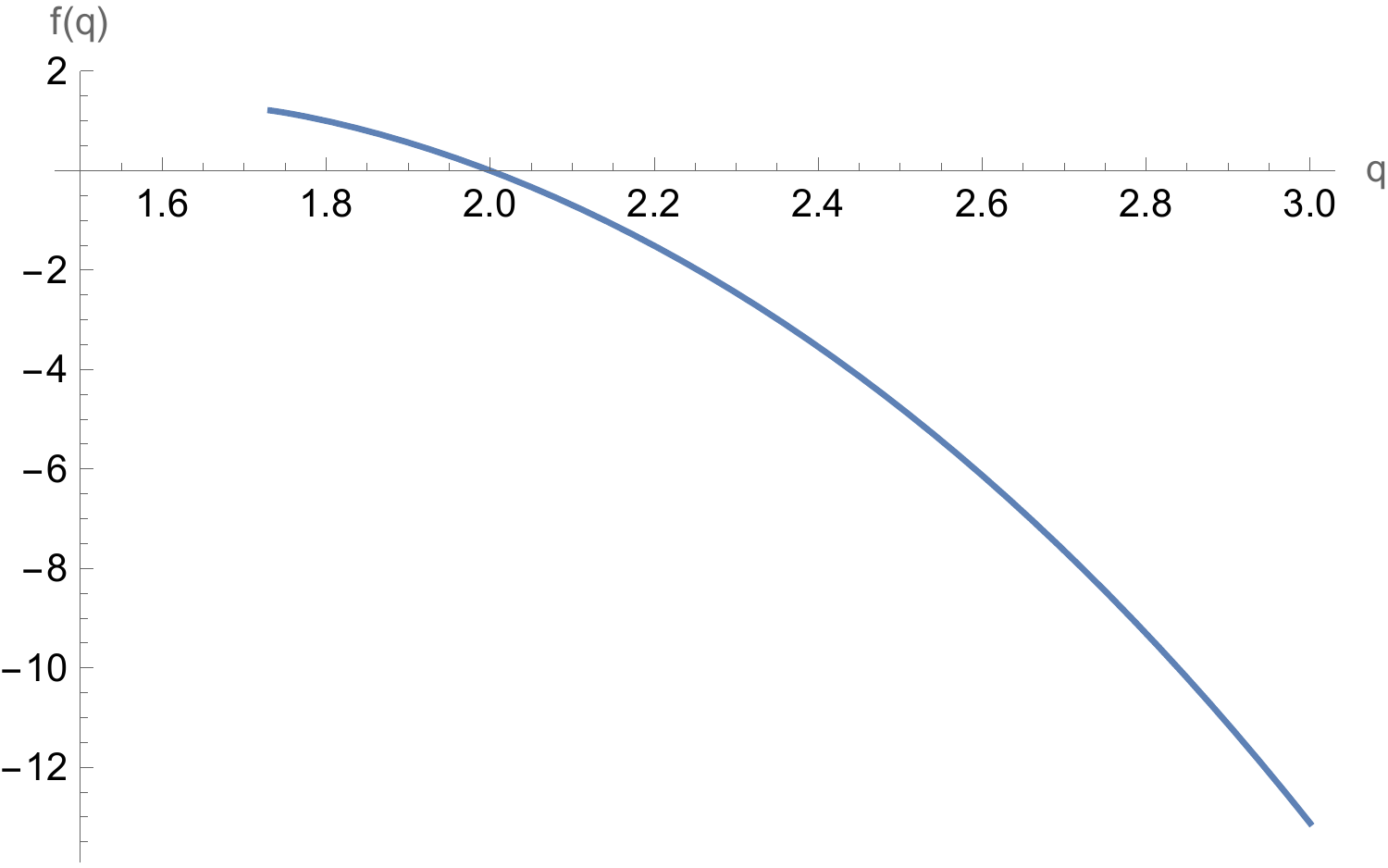}
\caption{
Auxiliary function $f(q)$ defined in \eqref{alpha_lambda_f}.
}
\label{fig_k_auxiliary_fun}
\end{figure}

As seen from \figref{fig_k_auxiliary_fun} energy density is a monotonically decreasing function of $v$, which is stretched in horizontal direction by $\lambda$ and in vertical direction by $\alpha \lambda^3$.
It is negative for $v > 2 \lambda$, so the ground state should have the smallest $\lambda$ of all other configurations.
Moreover, to have the lowest $\rho$ for all $v$ it should have the lowest asymptotic.
Namely, for $v \to +\infty$, we have $\rho \to - \frac{8 \pi}{27} \alpha v^3$, and hence the ground state should have the largest $\alpha$.

Firstly, let us consider the $\lambda$ parameter.
Configurations with one-sized discs ($N = 1$) have $\lambda = 1$ since $A_i = n_0$.
Let us prove that this is the lower bound for $\lambda$.
Namely, that
\begin{equation}\label{ineq_lambda}
\lambda^2 = \frac{A_{-1} A_1}{A_0^2} > 1\ \  \text{for}\ \ N > 1
\end{equation}

It is easy to show that \eqref{ineq_lambda} is true for $N = 2$.
Namely, that 
\begin{equation}
\frac{(n_0 + n_1 / \eta) (n_0 + \eta n_1)}{(n_0 + n_1)^2} > 1
\end{equation}
by rearranging we obtain
\begin{equation}
(\eta - 1)^2 n_0 n_1 / \eta > 0
\end{equation}
which is indeed true.

Next, we prove $N > 2$ cases of \eqref{ineq_lambda} by induction.
Assume that \eqref{ineq_lambda} is true for $N$ and let us prove it for $N + 1$.
Namely we need to prove that:
\begin{equation}
\frac{(A_{-1} + n_N / \eta_N) (A_1 + \eta_N n_N)}{(A_0 + n_N)^2} > 1\ \ \text{if}\ \ \frac{A_{-1} A_1}{A_0^2} > 1
\end{equation}
This inequality can be rearranged into the following form:
\begin{equation}
\frac{\left( \eta_N \sqrt{A_{-1}} - \sqrt{A_1} \right)^2}{2 \eta_N} + \sqrt{A_{-1} A_1} - A_0 + \frac{A_{-1} A_1 - A_0^2}{2 n_N} > 0
\end{equation}
which is true since the first term is a complete square, and the other terms are positive by assumption of induction.
This concludes the proof of \eqref{ineq_lambda}.

Now let us consider the $\alpha$ parameter.
For single-size discs $N = 1$ we obtain $\alpha = n_0 / S_0 = \delta / \pi$ where $\delta$ is density of discs -- area of discs divided by the total area.
The quantity $\delta$ is maximal for hexagonal packing of discs \figref{fig_hexagonal_packing} for which $\alpha_{max} = \frac{1}{2 \sqrt{3}}$.

Next, let us cover the plane by non-overlapping regions.
If all the regions indexed by $a$ have $\alpha^a \leq \alpha_{max}$ then the total $\alpha$ for the plane will be $\alpha \leq \alpha_{max}$.
To show that, consider two regions $a$ and $b$ with $\alpha^a \geq \alpha^b$.
Then we will prove that $\alpha^{ab} \leq \alpha^a$, namely:
\begin{equation}
\frac{\alpha^{ab}}{\alpha^a} = \frac{(1 + A)^3}{(1 + S) (1 + B)^2} \leq 1\ \  \text{for}\ \ \frac{\alpha^b}{\alpha^a} = \frac{A^3}{S B^2} \leq 1
\end{equation}
where $A = A_0^b / A_0^a$, $B = A_{-1}^b / A_{-1}^a$ and $S = S_0^b / S_0^a$.
Hence, we obtain that:
\begin{equation}
\frac{\alpha^{ab}}{\alpha^a} \leq \frac{(1 + (S B^2)^{1/3})^3}{(1 + S) (1 + B)^2} \leq 1
\end{equation}
where the last inequality can be rearranged into:
\begin{equation}
(B^{1/3} - S^{1/3})^2 (2 B^{1/3} + 2 B S^{1/3} + B^{4/3} + S^{1/3}) \geq 0
\end{equation}
which is indeed true.

\begin{figure}
\centering
\includegraphics[width=0.99\linewidth]{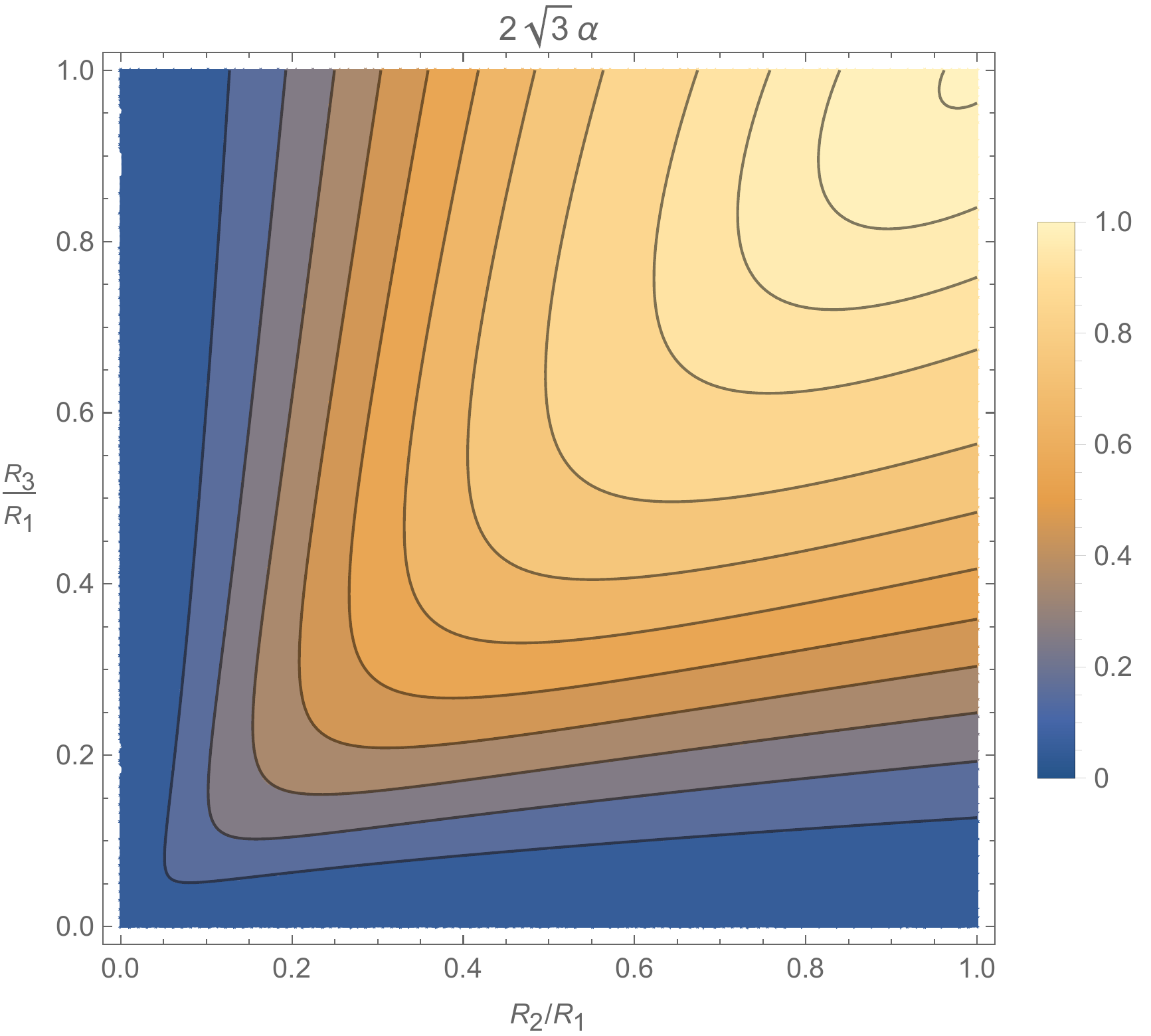}
\caption{
$\alpha$ parameter of a triangle formed by three touching discs as function of their radii.
It is maximal for $R_1 = R_2 = R_3$ and equals $\alpha = \alpha_{max} = \frac{1}{2 \sqrt{3}}$.
}
\label{fig_alpha_triangle}
\end{figure}

Now consider all the different compact packings of discs.
Contact graph of this type of packing is a triangulation of the plane.
By contact graph we mean a graph that has one vertex for each disc and an edge between pairs of discs that are mutually tangent.
Consider one of such triangles: it is formed by lines connecting the centers of three touching discs.
Without loss of generality assume that their radii are $R_1 \geq R_2 \geq R_3$.
We can then compute the $\alpha$ corresponding to this triangle, which attains its maximum $\alpha = \alpha_{max}$ when all radii are equal, see \figref{fig_alpha_triangle}.
Hence total alpha for the plane will be $\alpha \leq \alpha_{max}$ for compact packings of discs.

To summarize: we showed that among all the packings $\lambda = 1$ is the minimum, which is attained for all $N = 1$ packings.
Next, among all the compact packings $\alpha_{max}$ is the maximal value of $\alpha$, which is attained by hexagonal packing of discs.
These relations prove that hexagonal packing has energy lower than any compact disc packing.

\subsection{Proof that hexagonal packing has lower energy than any packing with lower density}
Let us show that for any disc packing:
\begin{equation}\label{alpha_bound_delta}
\alpha \leq \alpha_{max}\ \ \text{for}\ \ \delta < \delta_{hex} = \pi \alpha_{max}
\end{equation}

where $\alpha$ is defined in \eqref{alpha_lambda_f}, $\alpha_{max} = \frac{1}{2 \sqrt{3}}$ and density is $\delta = \pi A_2 / S_0$.
Using that and second inequality in \eqref{alpha_bound_delta} we obtain:
\begin{equation}
\alpha < \alpha_{max} \frac{A_0^3}{A_2 A_{-1}^2}
\end{equation}
Hence to prove first inequality in \eqref{alpha_bound_delta} it is sufficient to show that:
\begin{equation}\label{AAA_ineq}
\frac{A_0^3}{A_2 A_{-1}^2} \leq 1
\end{equation}
We will show that by induction.
For the case $N = 1$ when we have only one-sized discs we obtain $A_k = n_0$ and hence inequality \eqref{AAA_ineq} holds.
Then assuming that \eqref{AAA_ineq} is true for $N$ sizes of discs, let us show it for $N + 1$ size discs.
Namely, that:
\begin{equation}\label{alpha_bound_delta_Np1}
\frac{(A_0 + n_N)^3}{(A_2 + n_N \eta_N^2) (A_{-1} + n_N / \eta_N)^2} \leq 1
\end{equation}
if \eqref{alpha_bound_delta} is true.
Which is possible to show by rearranging \eqref{alpha_bound_delta_Np1} using \eqref{alpha_bound_delta}, that with $\widetilde{A} = A_2^{1/3} / (\eta_N A_{-1}^{1/3})$ and $\widetilde{n} = n_N / (\eta_N A_{-1})$ becomes:
\begin{equation}
- \left( \widetilde{A} - 1 \right)^2 \left( \widetilde{A} (2 + \widetilde{n}) + 1 + 2 \widetilde{n} \right) \leq 0
\end{equation}
which is indeed true.

So that proves the inequality \eqref{alpha_bound_delta} for $\alpha$.
In the previous subsection, we showed that among all the packings $\lambda = 1$ is the minimum.
Hence it shows that hexagonal packing has lower energy than any packing with lower density.

\section{Comparison of compact packings in the $R$ expansion model}\label{app_Rcomp}
\begin{figure}
\centering
\includegraphics[width=0.99\linewidth]{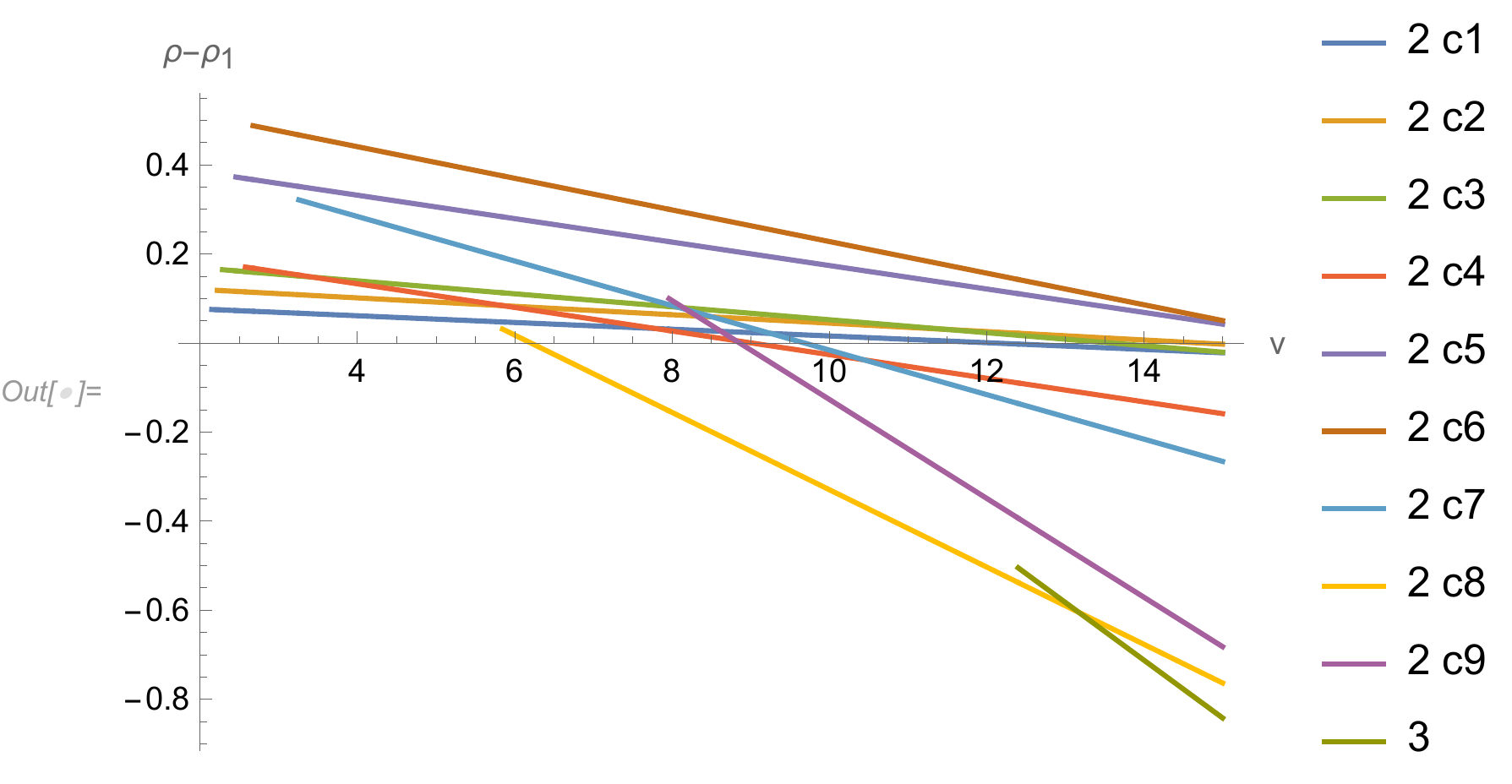}
\caption{
Difference of energy density of a given packing and a hexagonal packing \figref{fig_hexagonal_packing} as function of $v$ parameter in the model \eqref{gR} and \eqref{VR}.
The first number denotes the number of different sizes $N$ in the packing.
Packings with $N = 2$ are all possible compact packings \cite{kennedy2006compact}.
Packing with $N = 3$ is given in \figref{fig_hexagonal_packing_2}.
Simplest hexagonal packing \figref{fig_hexagonal_packing} has lowest energy for $2 < v < v_1$, where $v_1 = 6.20..$.
Packing $2\ c8$, which is a packing plotted in \figref{fig_hexagonal_packing_1}, is the ground state for $v_1 < v < v_2$, where $v_2 = 13.17..$.
For $v_2 < v$ Packing with $N = 3$, see \figref{fig_hexagonal_packing_2}, has lowest energy.
We expect packings with higher $N$ to become ground states for higher $v$.
}
\label{fig_packings_comparison_R}
\end{figure}

\begin{figure}
\centering
\includegraphics[width=0.99\linewidth]{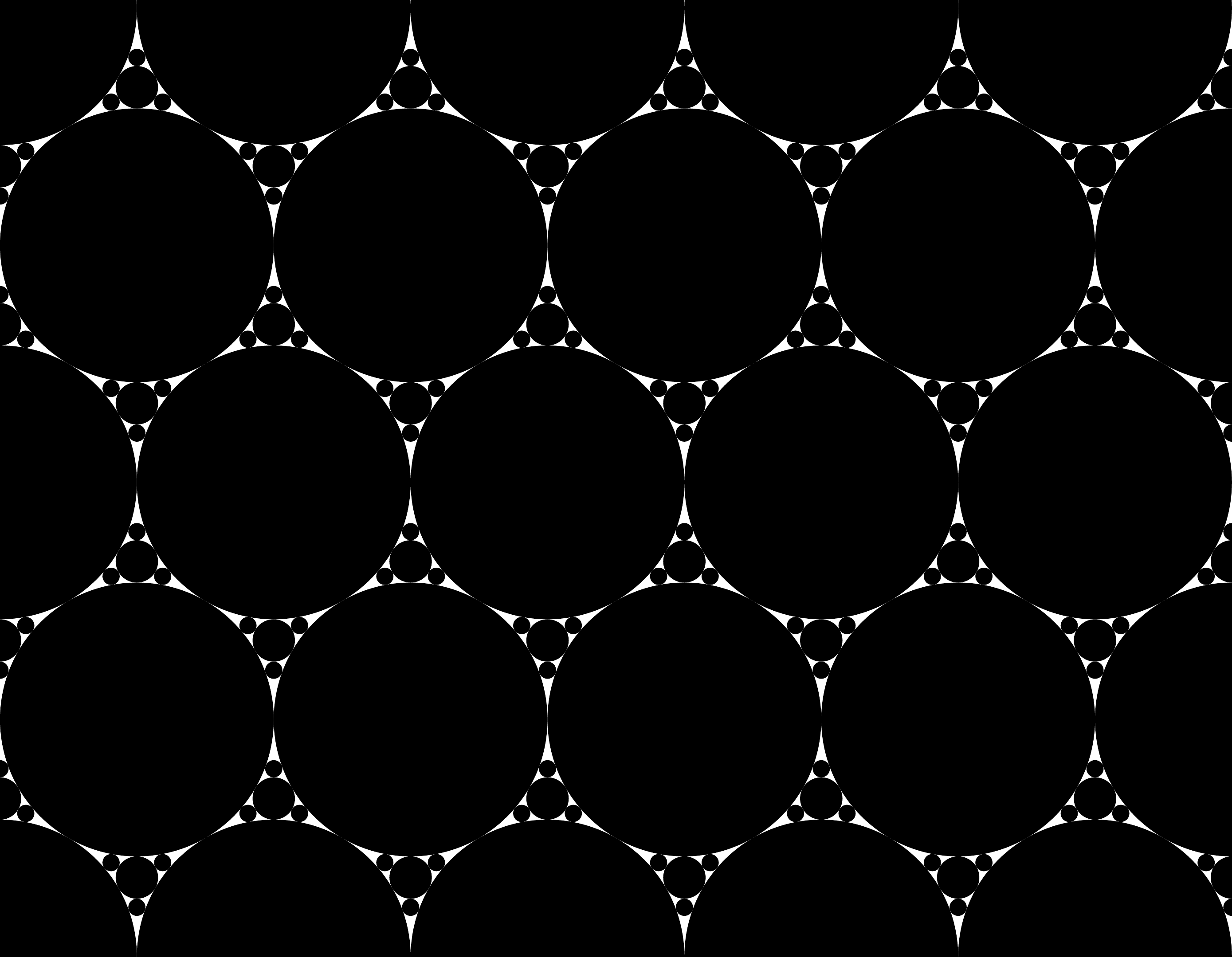}
\caption{
Hexagonal packing with three kinds of discs of different sizes.
}
\label{fig_hexagonal_packing_2}
\end{figure}

Here we consider model \eqref{VR}, which for any disc packing results in energy density:
\begin{equation}
\rho = 2 \pi \frac{A_1 R_0 - A_2 v R_0^2 + A_3 R_0^3}{S_0 R_0^2}
\end{equation}
minimizing it with respect to $R_0$ we obtain:
\begin{equation}
\begin{gathered}
R_0 = \sqrt{A_1 / A_3} \\
\rho =- 2 v \delta +  4 \pi \sqrt{A_1 A_3} / S_0
\end{gathered}
\end{equation}
where disc density $\delta = \pi A_2 / S_0$.
Which shows that when increasing $v$ configurations with lower density $\delta$ will become ground states.
We compare some compact packings in \figref{fig_packings_comparison_R}.
It shows that states \figref{fig_hexagonal_packing}, \figref{fig_hexagonal_packing_1}, \figref{fig_hexagonal_packing_2} are likely to be ground states as $v$ is increased.

Let us study the $v \to +\infty$ limit of this model.
By rescaling radii $R \to \widetilde{R} = v R$ and potential $V \to \widetilde{V} = V / v^2$ we obtain:
\begin{equation}
\widetilde{V}(\widetilde{R}) = \frac{1}{v^2} - \widetilde{R} + \widetilde{R}^2
\end{equation}
which means that for $v \to +\infty$ we get $\widetilde{V} \to - \widetilde{R} + \widetilde{R}^2$.
Note, that this model is unstable towards the formation of many infinitely small discs, as was shown in the main text.

\section{Other ways to have ground state fractals in two dimensional systems}\label{app_Other2dFractals}
To obtain fractal as a ground state we single out two conditions:

(i) We set some preferred \textit{size} for the mono-phase patch, such that energy $\to 0^-$ for size $\to 0$.
For example, one can have energy as a function of patch area $E(S)$, with the negative minimum at $S_0$ and $E \to - S^\alpha$ for $S \to 0$.
Note, that similar to the discussion in the main text we can set $\alpha > 1$ to have convergent solutions.

(ii) We fix the \textit{shape} of the interface between phases to the shape that cannot tile the plane.
This   ensures that there are gaps between large patches, which will be filled by smaller and smaller patches.
This can be done in many ways, so we discuss only a couple of examples, where the shape of the interface is a circle.

One way to fix the shape of the interface to a circle is to have energy that depends on the radius of the biggest circle that can be inscribed into the interface $R_{in}$ and the radius of the smallest circle in which the interface can be inscribed into $R_{out}$.
Then one can have energy $E(R_{out} / R_{in})$ such that $E(1) = 0$ and $E \to +\infty$ for $R_{out} / R_{in} \neq 1$.

\begin{figure}
\centering
\includegraphics[width=0.99\linewidth]{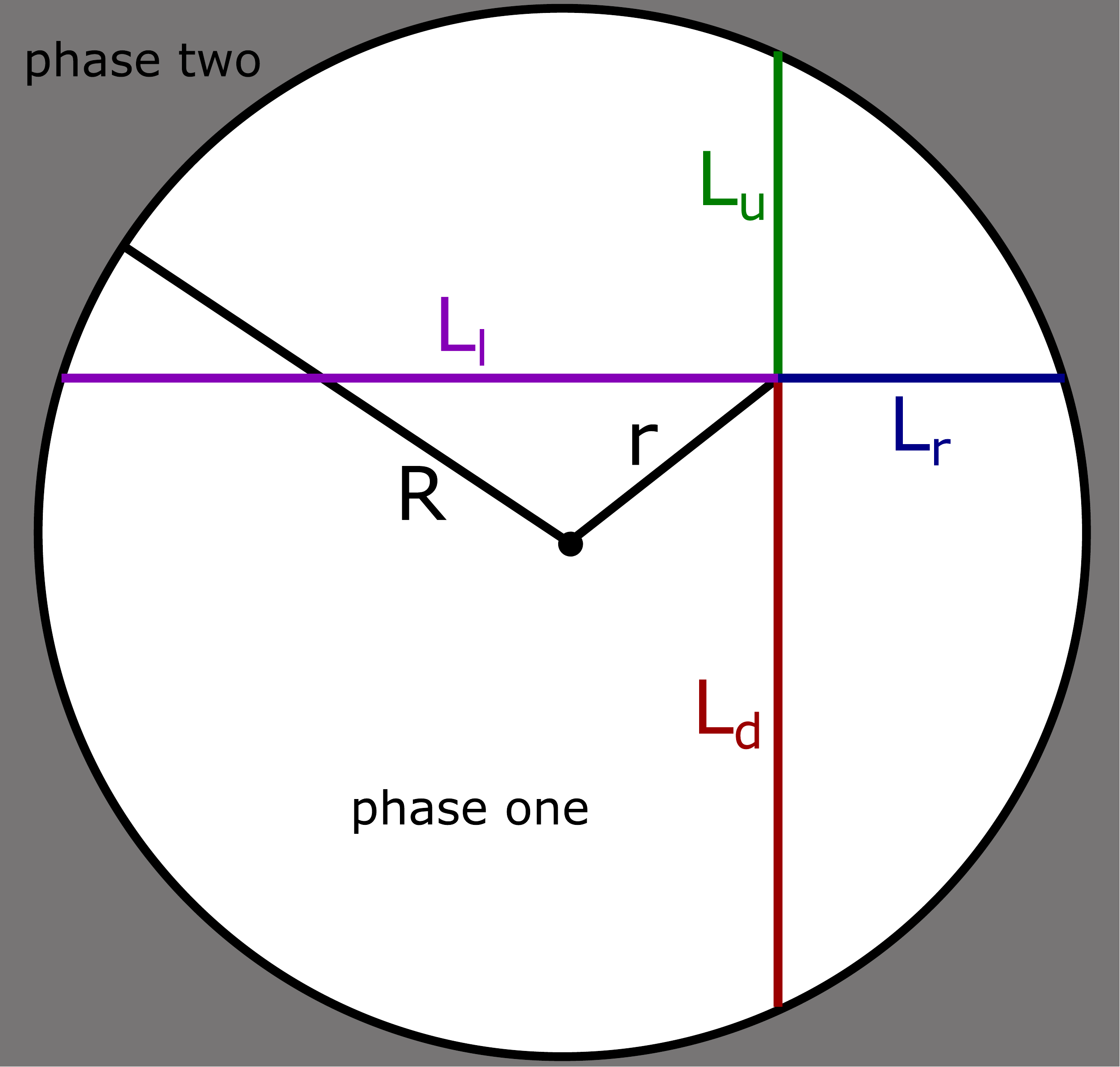}
\caption{
Illustration of distances to interface in up, down, left, right directions.
Note, that only in a circular interface $L_u L_d = L_l L_r = R^2 - r^2$.
}
\label{fig_udlr}
\end{figure}

Another way is to have energy density $\rho$ corresponding to every point in the plane, which depends on distances to the interface in four directions.
Namely, one can pick directions: up, down, left, right, see \figref{fig_udlr}.
Then energy density can be constructed such that $\rho = 0$ for $L_u L_d = L_l L_r$ and $\rho \to +\infty$ otherwise.
This will make interfaces circular.

Note, that in these and the example in the main text it is not necessary to fix the interface to exactly the circle to have a fractal ground state.
Namely, $\gamma$ parameter in \eqref{gk} can be finite or energy (density) for cases considered here can have some finite values instead of $\to +\infty$.
Then the interface will be non-circular, but still can be such that it will not tile the plane and hence will have fractal packing.

\section{Ground state fractal in one dimensional system}\label{app_1dFractals}
Here we consider a one-dimensional CSDF model with two phases that has fractal as the ground state.
It is defined on finite space of size $1$ and is given by energy:
\begin{equation}\label{H_1d}
H = \sum_{i = 1}^N E(L_i) + \sum_{i = 1}^{N - 1} C(L_{i+1} / L_i)
\end{equation}
where $N$ is number of line segments of alternating phases, $L_i$ is length of the $i$'s line segment, $E(L)$ is energy of a single line segment, $C(L_{i+1} / L_i)$ is energy of interface of the phases.
We choose energy $E$ such that short segments are preferred and set $E(L) = L^2$.
Energy of interface $C$ is chosen such that some given ratio of segment lengths $L_{i+1} / L_i = q$ is preferred.
Moreover, we consider the limit when:
\begin{equation}
\begin{gathered}
C(L_{i+1} / L_i) = 0\ \ \text{for}\ \ L_{i+1} / L_i = q\\
C(L_{i+1} / L_i) \to +\infty\ \ \text{for}\ \ L_{i+1} / L_i \neq q
\end{gathered}
\end{equation}

\begin{figure}
\centering
\includegraphics[width=0.99\linewidth]{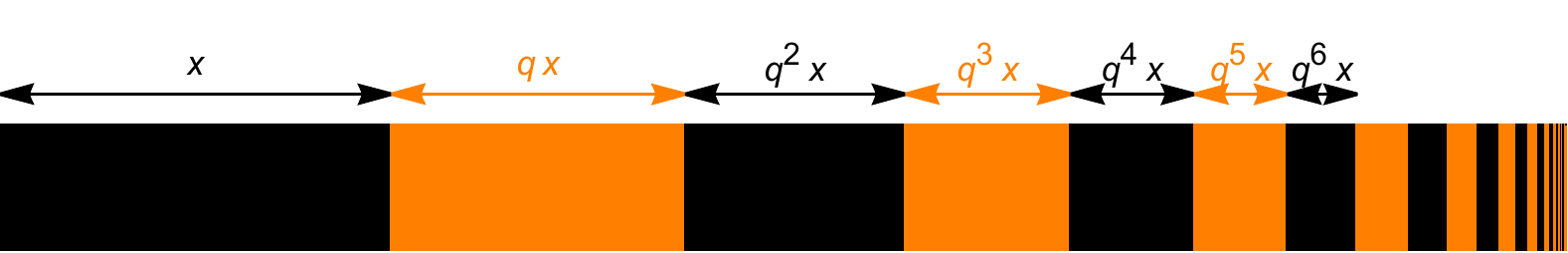}
\caption{
One dimensional fractal of alternating phase one (black) and phase two (orange) line segments.
Ratio of lengths of consecutive line segments is $q = 3/4$ and length of the longest segment is $x = 1 - q$.
}
\label{fig_1d_fractal}
\end{figure}

Hence the ratio is fixed to $L_{i+1} / L_i = q$.
In this case energy of $N$ line segments of lengths $L_i = x q^{i-1}$ equals $H = x^2 (1 - q^{2 N}) / (1 - q^2)$.
Using that total length is $\sum_{i = 1}^N L_i = 1$, we obtain $x$ and energy:
\begin{equation}
\begin{gathered}
x = \frac{1 - q}{1 - q^N} \to 1 - q\\
H = \frac{1 - q}{1 + q} \frac{1 + q^N}{1 - q^N} \to \frac{1 - q}{1 + q}
\end{gathered}
\end{equation}
where we take the limit $N \to +\infty$ since $H$ is a decreasing function of $N$.
Hence the ground state of the model \eqref{H_1d} is given by a fractal, see \figref{fig_1d_fractal}.

\end{document}